\documentclass[12pt]{article}
\usepackage[letterpaper,top=1in,bottom=1in,left=1in,right=1in]{geometry}

\usepackage{amsmath,amssymb}

\usepackage{pifont}

\usepackage{changepage}

\usepackage[utf8x]{inputenc}

\usepackage{textcomp,marvosym}

\usepackage{nameref,hyperref}

\usepackage{booktabs}

\usepackage{placeins}

\usepackage[right]{lineno}

\usepackage{microtype}
\DisableLigatures[f]{encoding = *, family = * }

\usepackage[table]{xcolor}

\usepackage{subcaption}

\usepackage{array}
\usepackage{blkarray}
\usepackage{multirow}
\usepackage{mathtools}

\usepackage{xr-hyper}
\usepackage[capitalise]{cleveref}

\usepackage{tikz}
\usetikzlibrary{matrix}

\usepackage[aboveskip=1pt,labelfont=bf,labelsep=period]{caption}

\usepackage{authblk}

\usepackage{setspace}
\renewcommand{\textapprox}{\raisebox{0.5ex}{\texttildelow}}

\crefname{figure}{Figure}{Figures}
\crefname{table}{Table}{Tables}

\title{SynCoTrain: A Dual Classifier PU-learning Framework for Synthesizability Prediction}

\author[1]{Sasan Amariamir}
\author[1,2*]{Janine George}
\author[1\textdagger]{Philipp Benner}
\affil[1]{Federal Institute of Materials Research and Testing, Unter den Eichen 87, 12205 Berlin, Germany}
\affil[2]{Friedrich Schiller University Jena, Institute of Condensed Matter Theory and Solid-State Optics, Max-Wien-Platz 1, 07743 Jena (Germany)}
\affil[*]{\textit{janine.george@bam.de}, {}\textsuperscript{\textdagger}\textit{philipp.benner@bam.de}}

\begin{document}

\maketitle

\begin{abstract}
Material discovery is a cornerstone of modern science, driving advancements in diverse disciplines from biomedical technology to climate solutions. Predicting synthesizability, a critical factor in realizing novel materials, remains a complex challenge due to the limitations of traditional heuristics and thermodynamic proxies. While stability metrics such as formation energy offer partial insights, they fail to account for kinetic factors and technological constraints that influence synthesis outcomes. These challenges are further compounded by the scarcity of negative data, as failed synthesis attempts are often unpublished or context-specific.

We present SynCoTrain, a semi-supervised machine learning model designed to predict the synthesizability of materials. SynCoTrain employs a co-training framework leveraging two complementary graph convolutional neural networks: SchNet and ALIGNN. By iteratively exchanging predictions between classifiers, SynCoTrain mitigates model bias and enhances generalizability. Our approach uses Positive and Unlabeled (PU) Learning to address the absence of explicit negative data, iteratively refining predictions through collaborative learning.

The model demonstrates robust performance, achieving high recall on internal and leave-out test sets. By focusing on oxide crystals, a well-characterized material family with extensive experimental data, we establish SynCoTrain as a reliable tool for predicting synthesizability while balancing dataset variability and computational efficiency. This work highlights the potential of co-training to advance high-throughput materials discovery and generative research, offering a scalable solution to the challenge of synthesizability prediction.
\end{abstract}

\section{Introduction}

Material discovery is a foundational pillar of modern science and
perhaps the driving motivation behind materials science. It supports
advancements in numerous scientific and technological disciplines. In
this field, the ability to predict synthesizability is crucial.
Developing materials with novel properties expands the possibilities in
endeavors from functional materials used in biomedical devices to
addressing the challenges of climate change~\cite{de_pablo_new_2019}. In the past decade
or so, efforts such as the Materials Genome Initiative aimed to
accelerate the discovery, development, and deployment of new materials in
the hopes of societal betterment~\cite{de_pablo_new_2019, white_materials_2012}. An essential part of
realizing this goal is employing high-throughput simulations and
experiments for screening candidate materials with desirable
properties~\cite{de_pablo_new_2019, rodgers_materials_2006}. Unfortunately, a substantial amount of
resources and effort can be wasted on hypothetical materials that
currently cannot be synthesized.

Historically, physico-chemical based heuristics such as the Pauling
Rules~\cite{pauling_principles_1929} or the charge-balancing criteria~\cite{antoniuk_predicting_2023} have been used to
assess materials stability and synthesizability. Nevertheless, these
simplified approaches have been shown to be out of date, as more than
half of the experimental (already synthesized) materials on the
Materials Project database~\cite{jain_commentary_2013} do not meet these criteria for
synthesizability~\cite{antoniuk_predicting_2023, george_limited_2020}.

In more recent attempts, material scientists often employed
thermodynamic stability as a proxy for synthesizability, ignoring the
effect of kinetic stabilization. This involves conducting
first-principle calculations to estimate the formation energy of
crystals and their distance from the convex hull. A negative formation
energy, or a minimal distance from the convex hull, is commonly
interpreted as an indicator of synthesizability~\cite{sun_thermodynamic_2016,cerqueira_identification_2015,singh_robust_2019,
noh_inverse_2019, wu_first_2012}.
While stability significantly contributes to synthesizability, it is
just one aspect of this complex issue. There are many -potentially
interesting- metastable materials that do exist, even though their
formation energies deviate from the ground-state~\cite{sun_thermodynamic_2016, noh_inverse_2019, bartel_review_2022, jang_structure-based_2020, lee_machine_2022}. These materials can be synthesized in
alternate thermodynamic conditions in which they are the ground-state.
After removing the favorable thermodynamic field, they have stayed stuck
in the metastable structure by kinetic stabilization~\cite{sun_thermodynamic_2016}. On the
other hand, there are many hypothetical stable materials in
well-explored chemical spaces which have never been synthesized. This
could be due to a high activation energy barrier between them and the
common precursors~\cite{bartel_review_2022, jang_structure-based_2020, lee_machine_2022}. Beyond the theoretical
and thermodynamic considerations, synthesizability is also a
technological problem. Novel materials that are developed through
cutting-edge methods were practically unsynthesizable before the
invention of their methods of synthesis. For example, new high-entropy alloys with great potential for catalysis applications were recently synthesized using the Carbothermal Shock (CTS) method~\cite{li_recent_2021}. Their particular homogeneous components and uniform structures were not accessible through conventional synthesis methods.
On the other hand, some
materials can only be synthesized under specific conditions, such as
extremely high pressures~\cite{miao_chemistry_2020} or within solvents like liquid
ammonia~\cite{liu_plastic_2020}, which alter their chemical behavior. Once these
conditions are removed, however, the materials may decompose.

The fact that estimating synthesizability is related to materials
structures without a straightforward formula to solve for, makes it an
apt candidate for machine learning. We define a classification task for
two classes of materials, namely synthesizable (the positive class), and
unsynthesizable (the negative class). This classification comes with a
few challenges and intricacies. The first one is encoding materials
structures in a machine understandable format. Some previous works have
circumvented this challenge in creative ways such as combining different
elemental features~\cite{lee_machine_2022, legrain_materials_2018}, using text-mining algorithms to
search the relevant literature to identify synthesizable
materials~\cite{huo_semi-supervised_2019}, using the picture of crystal cells with convolutional
neural networks~\cite{davariashtiyani_predicting_2021}, or even a network analysis of materials
discovery timeline with respect to their stability~\cite{aykol_network_2019}.
Others~\cite{jang_structure-based_2020,gu_perovskite_2022}, including this work, utilize graph
convolutional neural networks (GCNNs) to encode and learn from crystal
structures. While the GCNNs are more complicated to implement, they have
the advantage of including more information about the structure than composition alone or the other previously mentioned approaches that represent the structure information
indirectly through a proxy.

The second challenge of estimating synthesizability lies within the
nature of the available data. Unlike a typical classification task, we
do not have access to enough negative data. On the one hand, this is due
to the fact that unsuccessful attempts of synthesis are not typically
published nor uploaded to public databases. The attempts of using such
failed experiments~\cite{raccuglia_machine-learning-assisted_2016} inevitably remain confined to local labs and
a small class of materials. Also, synthesis success strongly depends on
the synthesis conditions and technology. Hence, the failure of synthesis
attempts in one setting does not necessarily imply failure in a
different lab with different synthesis methods or equipment. Finally,
creating a proper negative-set for training a classifier is a whole new
challenge~\cite{antoniuk_predicting_2023}. If the negative-set is too different from real
materials, it may not teach the model a meaningful decision boundary for
detecting synthesizability. To design a realistic-looking negative-set,
one would need to understand the features that determine
synthesizability in the first place.

The final challenge in this task comes as a fundamental aspect of
machine learning. Regardless of which model is chosen, it will
inherently exhibit a certain degree of bias. One introduces an
unintended bias when selecting one model over another, since the model's
ability to generalize out of sample is, in part, predetermined by its
architecture. This model bias comes even with the best performing
models. In fact, a model with great benchmarks might perform worse than
simpler models when predicting targets for out-of-distribution
data~\cite{li_critical_2023}, perhaps due to overfitting. This challenge becomes
particularly pronounced when predicting synthesizability. The objective
is to forecast a target for new and often out-of-distribution data,
where the issue of generalization is most acute. The lack of the
negative data compounds this issue, as it makes the accuracy or recall
less reliable. One way to mitigate this inherent bias is to leverage
multiple models~\cite{li_critical_2023}.

To address these challenges, we have developed a model ready for
integration into high-throughput simulations and generative materials
research. It is called \emph{SynCoTrain} (pronounced similar to
'Synchrotron'). It is a semi-supervised classification model designed
for predicting synthesizability of oxide crystals. \emph{SynCoTrain}
addresses the generalizability issue by utilizing co-training.
Co-training is an iterative semi-supervised learning process designed
for scenarios with some positive data and a lot of unlabeled
data~\cite{blum_combining_1998, denis_text_2003}. It leverages the predictive power of two
distinct classifiers to find and label positive data points among the
unlabeled data. Different models have different biases, and by combining
their predictions, we can practically reduce these biases while keeping
what they learn about the target. We use the Atomistic Line Graph Neural
Network (ALIGNN)~\cite{choudhary_atomistic_2021} and the SchNetPack~\cite{schutt_schnetpack_2019, schutt_schnet_2017} models as
our chosen classifiers. They are both innovative GCNNs with distinct
attributes. ALIGNN is unique in that it directly encodes both atomic bonds and bond angles into its architecture, offering a perspective that aligns with a chemist's view of the data. SchNetPack stands out for using a
unique continuous convolution filter which is suitable for encoding
atomic structures, which can be thought of as a physicist's perspective on the data.

At each step of co-training, \emph{SynCoTrain} learns the distribution
of the synthesizable crystals through the Positive and Unlabeled
Learning (PU Learning) method introduced by Mordelet and Vert~\cite{mordelet_bagging_2010}.
This base PU Learning method with a different classifier has already
been employed to predict synthesizability for all classes of
crystals~\cite{jang_structure-based_2020} and for perovskites specifically~\cite{gu_perovskite_2022}. In this work,
we utilize multiple PU Learners as the building blocks for co-training.
In each iteration of co-training, the learning agents exchange the
knowledge they gained from the data between each other. Eventually, the
labels are decided based on average of their predictions. This process
increases the prediction reliability and accuracy, much like two experts
who discuss and reconcile their views before finalizing a complex
decision. This collaborative approach suggests that co-training is more
likely to generalize effectively to unseen data compared to using a
single model with equivalent classification metrics such as accuracy or
recall.

We verify the performance of the model by recall for an internal
test-set and a leave-out test-set. We also evaluate our model further by
predicting whether a crystal is stable or not for the same data points.
Note that in predicting stability, we do not aim for a good performance.
In fact, we expect an overall poor performance due to high contamination
of the unlabeled data~\cite{mordelet_bagging_2010}; more info in supplemental material.
However, we compare the ground truth recall in stability to the recall
produced by the PU Learning, to gauge the reliability of the latter.

We chose a single family of materials, oxides, to establish the utility
of co-training in predicting materials properties. Oxides are a
well-studied class of materials with a large amount of experimental data
to learn from~\cite{savkina_oxide-based_2020, waroquiers_statistical_2017}. A higher number of training data would
typically decrease the classification error in machine learning.
However, training across all available families of crystals would
introduce greater variability in the dataset, potentially increasing the
uncertainty and error margins in our results. In other words, the
prediction quality for new materials would vary substantially. By
achieving high recall values with oxides as our training data, we
demonstrate the effectiveness of co-training. This approach ensures
reliable results while maintaining reasonable training times for our
models.

\hypertarget{result-and-discussion}{%
\section{Results and Discussion}\label{result-and-discussion}}

\hypertarget{model-development}{%
\subsection{Model Development}\label{model-development}}

\begin{figure}[!htbp]
    \centering
    \begin{subfigure}[b]{0.49\textwidth}
        \centering
        \includegraphics[height=4in]{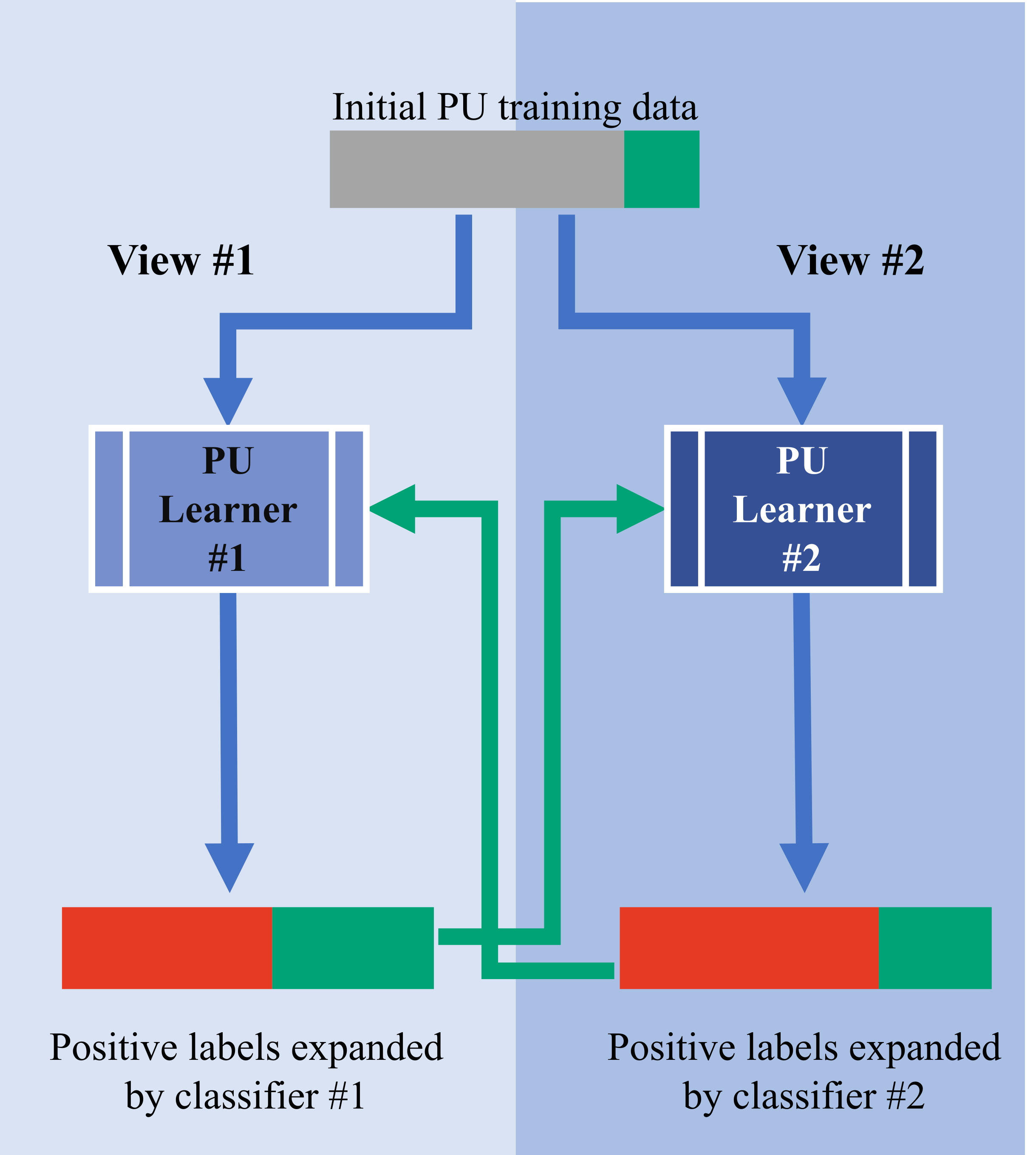}  
        \caption{}
        \label{fig:part-a}
    \end{subfigure}
    \begin{subfigure}[b]{0.49\textwidth}
        \centering
        \includegraphics[height=3.7in]{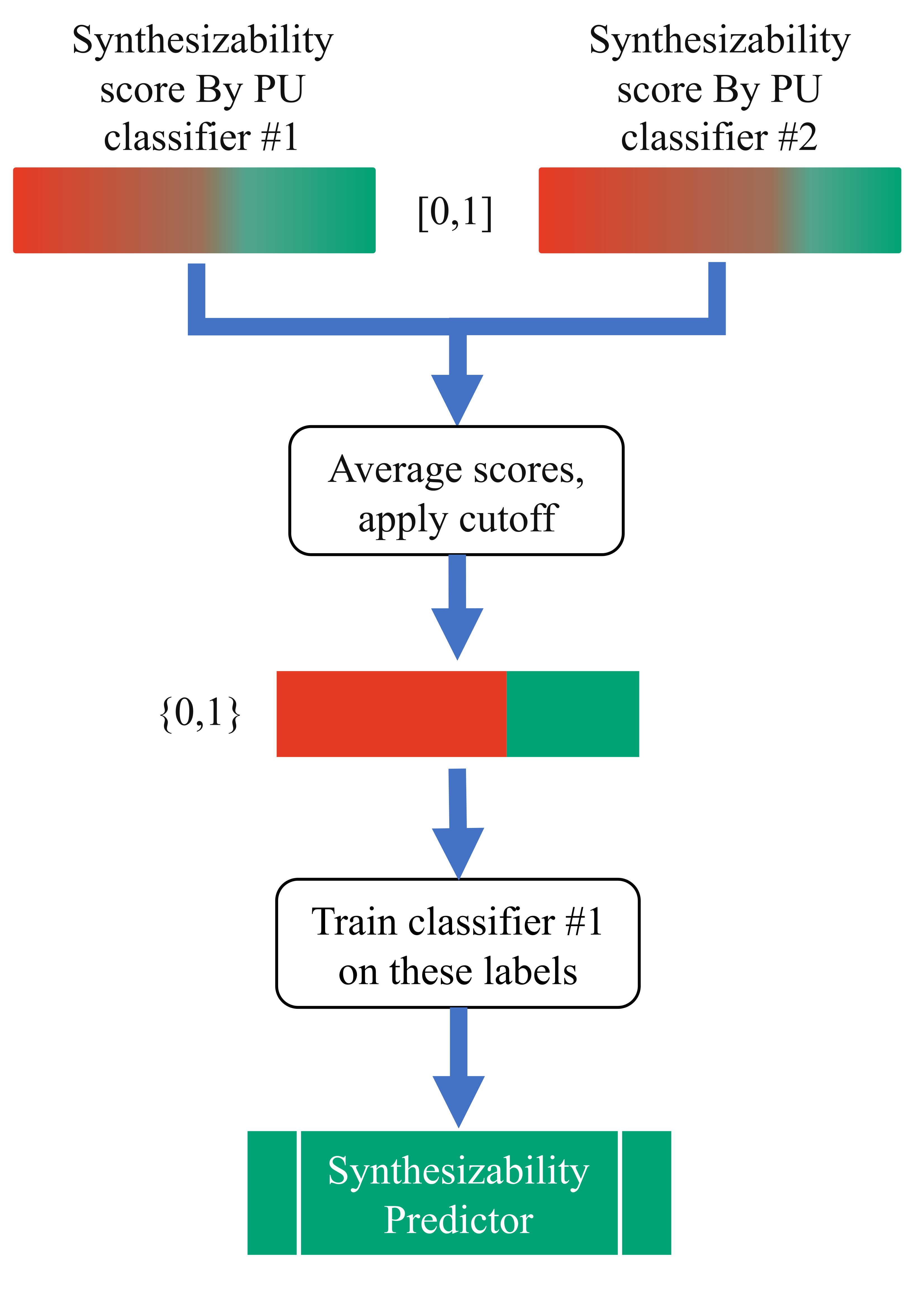}  
        \caption{}
        \label{fig:part-b}
    \end{subfigure}
    \begin{subfigure}[b]{0.98\textwidth}
        \centering
        \includegraphics[width=\textwidth]{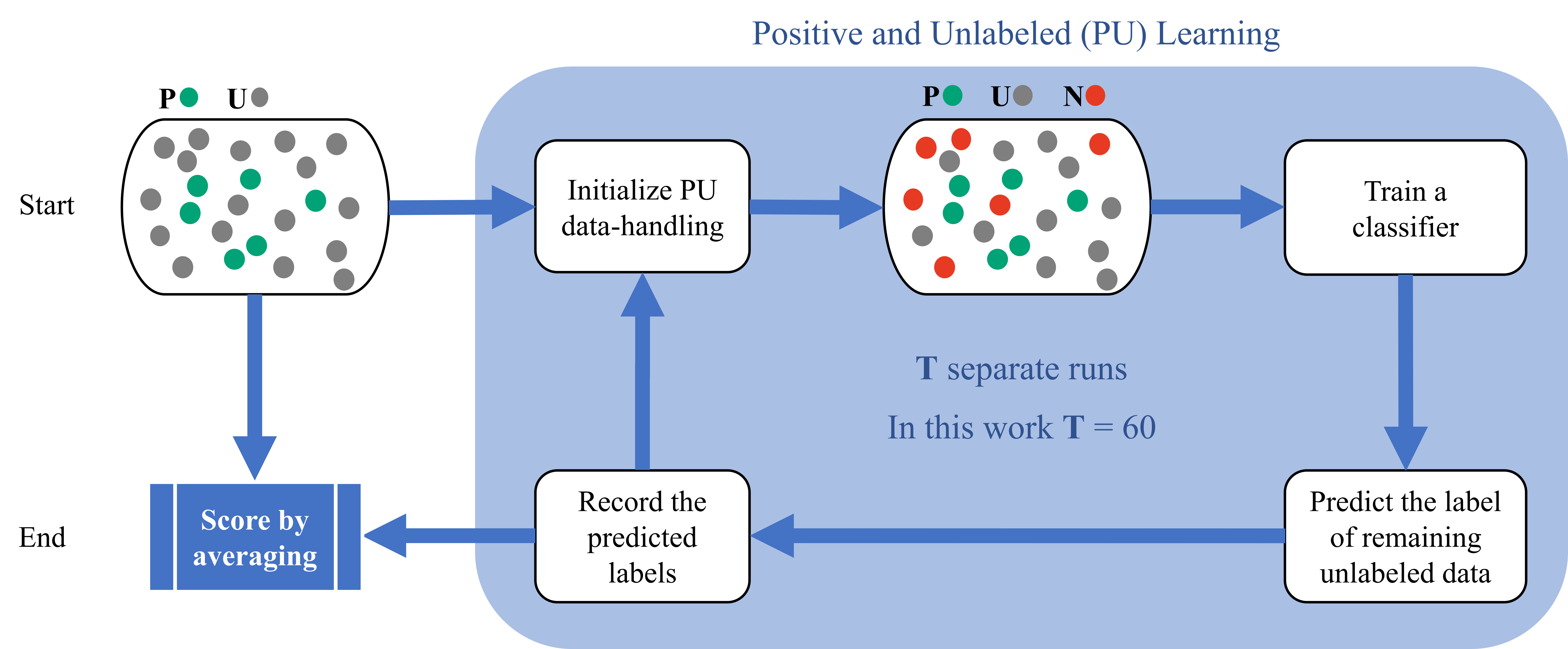}
        \caption{}
        \label{fig:part-c}
    \end{subfigure}
    
    \caption{\textit{Overview of the Workflow in SynCoTrain \textbf{a)} The PU data is passed to two distinct PU classifiers, each learning from a different view of the data. Each classifier labels unlabeled data points as positive or negative. The new labels from each PU classifier are used to expand the positive class for retraining the other classifier. \textbf{b)} After co-training steps, each unlabeled data point receives a prediction score from each PU classifier. An average of these scores is calculated for each data point, and a cutoff is applied to produce a label. All the data, now labeled, are used to train a final classifier to predict synthesizability. \textbf{c)} The PU learning process. Positive, negative, and unlabeled data are depicted as green, red, and gray circles respectively. Each run starts with training a classifier, with a randomly chosen subset of the unlabeled data used as the negative class. Labels are predicted for the remaining unlabeled data, and final scores are computed by averaging these predictions.}}
    \label{fig:model-development}
\end{figure}

The data for oxide crystals were obtained from the \textbf{Inorganic
Crystal Structure Database (ICSD)}~\cite{Hellenbrandt_icsd_2004}, accessed through the
\textbf{Materials Project API}~\cite{jain_commentary_2013}. The experimental and theoretical
data are distinguished based on the `theoretical' attribute. We used the
get\_valences function of \textbf{pymatgen}~\cite{ong_python_2013} to include only
oxides where the oxidation number is determinable and the oxidation
state of oxygen is -2.

Less than 1\% of the experimental data with energy above hull higher
than 1eV were removed, as potentially corrupt data. The learning began
with 10206 experimental and 31245 unlabeled data points.

Co-training consists of two separate iteration series, the results of
which are averaged in the final step. In the first series, we start by
training a base PU learner with an ALIGNN classifier. This is the
iteration `0' of co-training, and this step is called ALIGNN0. The
learning agent predicts positive labels for some of the unlabeled data,
creating a pseudo-positive class. This class is added to the original
experimental data, expanding the initial positive class. Iteration `1'
of co-training on this series is to train a base PU learner with the
other classifier, here the SchNet, on the newly expanded labels. This
step is called coSchnet1. Each iteration provides newly expanded labels
for the next iteration. The classifiers alternate for each iteration,
from ALIGNN to SchNet and vice versa, as shown in Fig.~\ref{fig:part-a}.

Parallel to this series, we set up a mirror series where iteration `0'
begins with a SchNet based PU learner. This step of iteration `0' is
called SchNet0. This series learns the data from a different,
complementary view compared to the former series, see Fig.~\ref{fig:part-a}.
It
continues in the same manner with alternating classifiers. The order of
the steps in each series can be found in \emph{Table~\ref{tab:cotraining_steps}}.

Each base PU learner produces a synthesizability score between 0 and 1
for each unlabeled datum. This is done through 60 runs of the bagging
method established by Mordelet and Vert~\cite{mordelet_bagging_2010}, as illustrated in Fig.~\ref{fig:part-c}.
In each independent run of this ensemble learner, a random subset of
the unlabeled data is sampled to play the role of the negative data in
training the classifier. The average of the predictions in these runs
for data points that were not part of the training in that run yields
the synthesizability score. This score is interpreted as the predicted
probability of being synthesizable. A threshold of 0.5 is applied for
labeling each datum as either synthesizable (labeled 1) or
not-synthesizable (labeled 0).

After several iterations of co-training, the optimal iteration is chosen
based on the prediction metrics. Continuing to further iterations yields
diminishing returns in performance metric while risking reinforcing
existing model bias. The scores provided from the two series at the
optimal iteration are then averaged. The 0.5 cutoff threshold is applied
to this averaged score to produce the final synthesizability score. Once
we have synthesizability labels for both the experimental and
theoretical data, a simple machine learning task remains. We train a
classifier on these labels and end up with a model that can predict
synthesizability (see Fig.~\ref{fig:part-b}).

\begin{table}[!htbp]
    \centering
    \caption{Co-training steps}
    \label{tab:cotraining_steps}
    \resizebox{\textwidth}{!}{%
    \begin{tabular}{@{}l !{\vrule width 2pt} >{\centering\arraybackslash}p{0.13\textwidth} c  > {\centering\arraybackslash}p{0.13\textwidth} c >{\centering\arraybackslash}p{0.13\textwidth} c >{\centering\arraybackslash}p{0.13\textwidth} !{\vrule width 2pt} >{\centering\arraybackslash}p{0.24\textwidth} @{}}
    \toprule
    \midrule
    \textbf{Co-training steps} & \textbf{Iteration `0'} & & \textbf{Iteration `1'} & & \textbf{Iteration `2'} & & \textbf{Iteration `3'} & \textbf{Averaging scores} \\
    
    \midrule
    \textbf{Training data source} & Original\newline labels & & Labels\newline expanded by\newline Iteration `0' & & Labels\newline expanded by\newline Iteration `1' & & Labels\newline expanded by\newline Iteration `2' & Scores provided\newline by the optimal\newline iteration \\
    \cmidrule(r){1-9}  
    
    \textbf{Training series} & ALIGNN0 & \textgreater{} & coSchnet1 & \textgreater{} & coAlignn2 & \textgreater{} & coSchnet3 & Synthesizability scores \\
    \cmidrule(r){2-8}  
    
    & SchNet0 & \textgreater{} & coAlignn1 & \textgreater{} & coSchnet2 & \textgreater{} & coAlignn3 & \\
    \bottomrule
    \end{tabular}%
    }
\end{table}

\subsection{Model Evaluation and Results}
\label{model-evaluation-and-results}
\begin{figure}[!htbp]
    \centering
    \begin{subfigure}[b]{0.6\textwidth}  
        \centering
        \includegraphics[width=\textwidth]{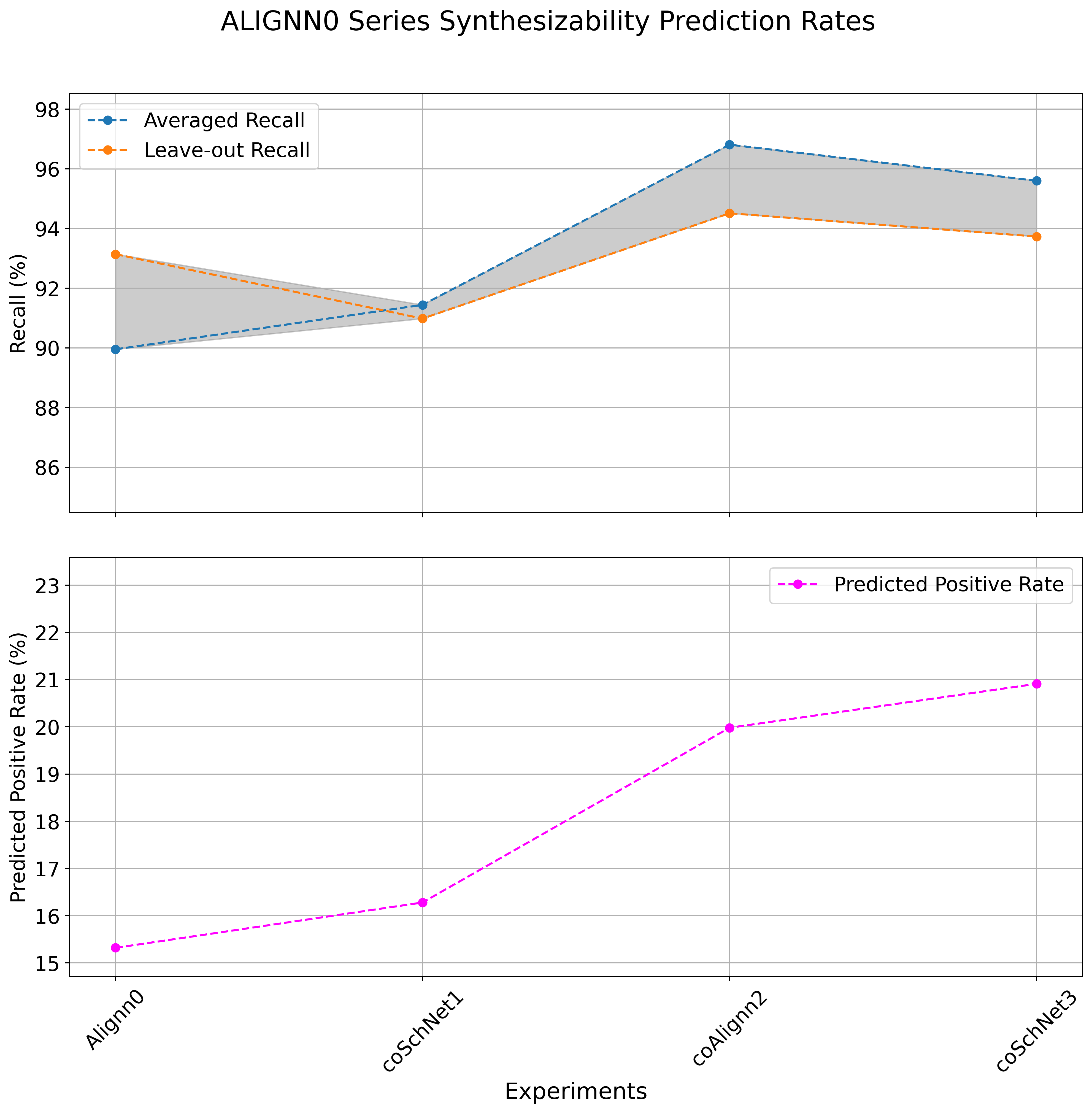} 
        \caption{}
        \label{fig:part-2a}
    \end{subfigure}
    
    \begin{subfigure}[b]{0.6\textwidth}  
        \centering
        \includegraphics[width=\textwidth]{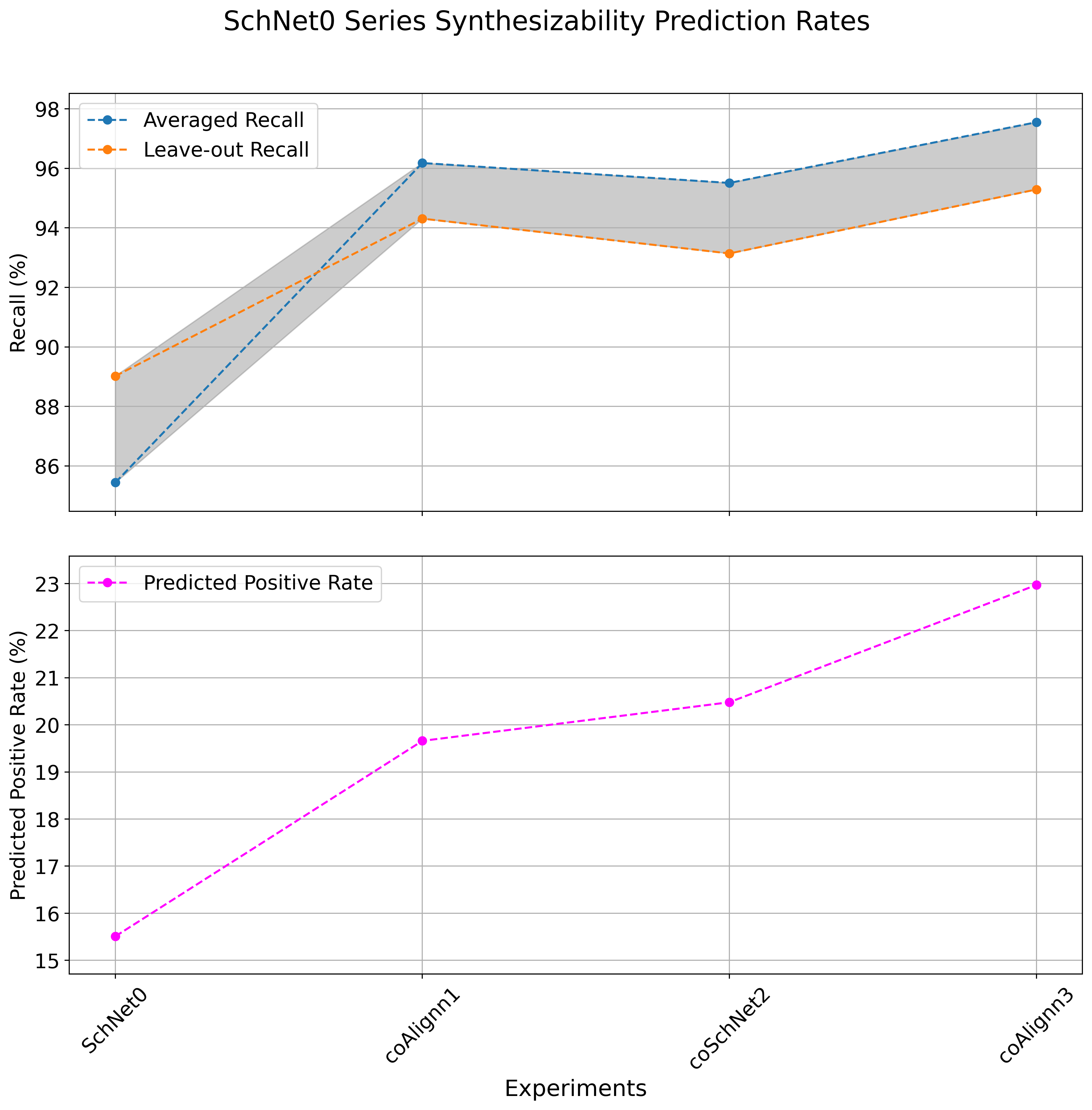}  
        \caption{}
        \label{fig:part-2b}
    \end{subfigure}
    
    \caption{\textit{Recall progression per iteration for both co-training series}}
    \label{fig:recallRate}
\end{figure}

Within each single run of the base PU Learning, the classifiers optimize
for Accuracy, as they operate unaware of the PU nature of the data. When
reporting the performance of PU learning, however, one should not use
Accuracy, Precision or the F1-score. These common measures assume
knowledge of the negative labels and a False Positive count. We use
Recall, also known as Sensitivity or True Positive Rate (true
positive/(true positive + false negative)) to report and benchmark the
performance of our PU learner as it only relies on the knowledge of the
positive data.

In our study, we employ two distinct test-sets to measure Recall. The
first is a dynamic test-set, which varies with each iteration of base PU
learning. The second is a leave-out test-set that remains untouched
during all training iterations. As the result, we obtain a `recall
range' between the two distinct Recall measures; an averaged recall for
the dynamic test-set and a leave-out recall. This gives us more
information than a single recall value. The construction and reasoning
behind this are detailed in the Ground Truth evaluation
section.

The construction and reasoning behind this are detailed in the Ground Truth evaluation section.

\begin{figure}[!htbp]
    \centering
    \begin{subfigure}[b]{0.45\textwidth}
        \centering
        \includegraphics[width=\textwidth]{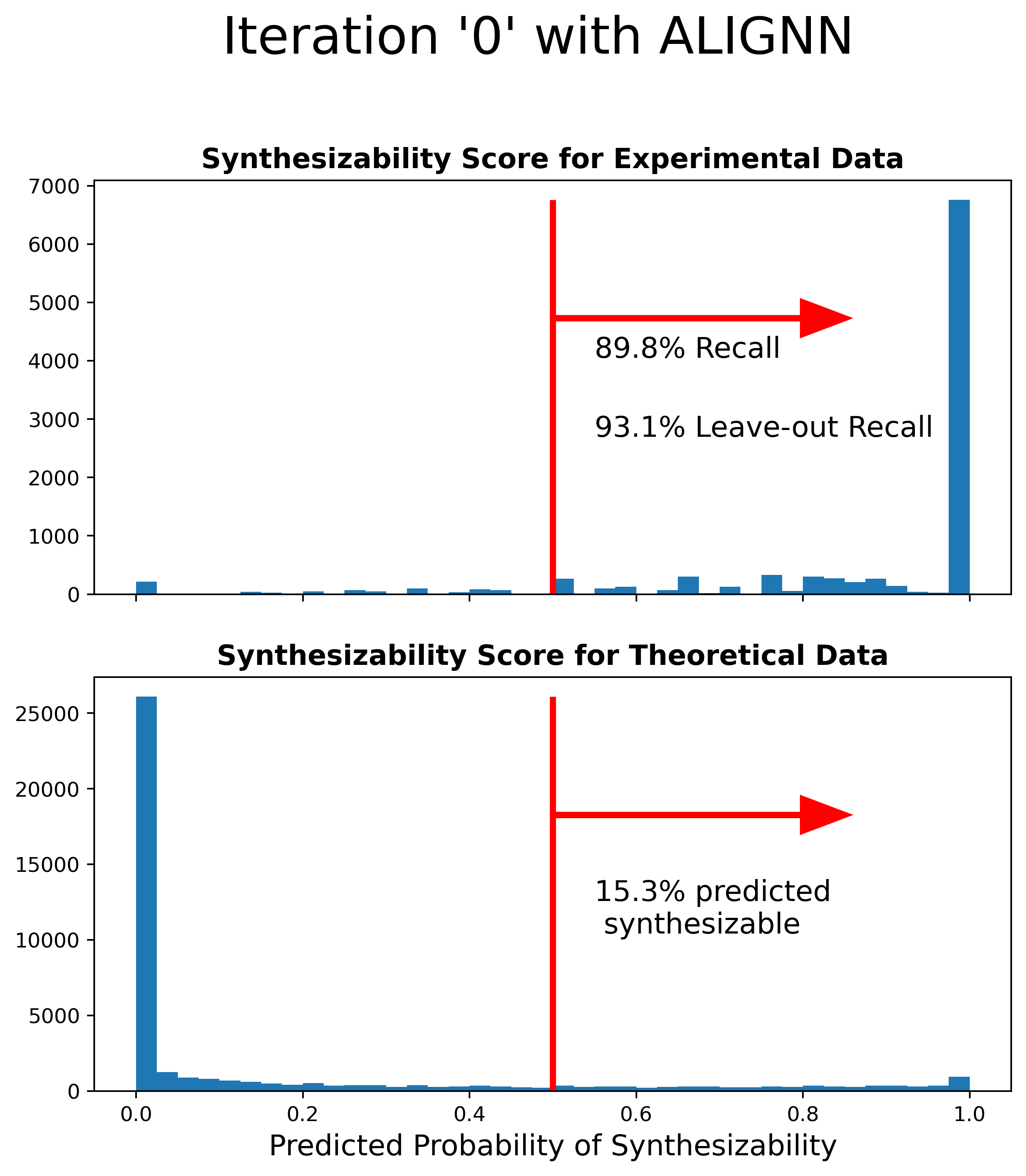}
        \caption{}
    \end{subfigure}
    \hfill
    \begin{subfigure}[b]{0.45\textwidth}
        \centering
        \includegraphics[width=\textwidth]{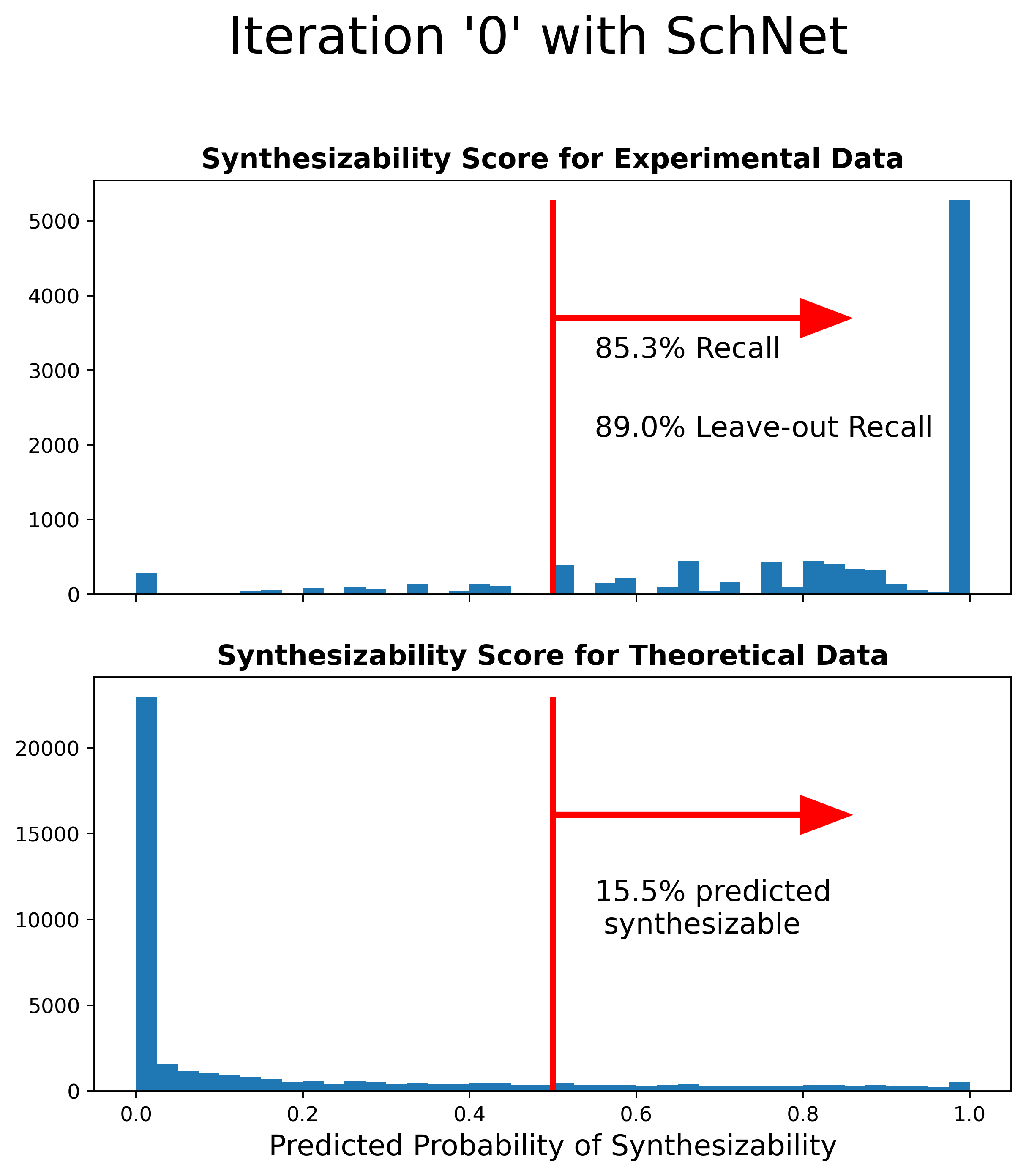}
        \caption{}
    \end{subfigure}
    \caption{\textit{Synthesizability score distribution for Iteration '0' for the a) ALIGNN0 series and b) SchNet0 series.}}
    \label{fig:synthscoredist_0}
\end{figure}

The recall values for each iteration are depicted in Fig.~\ref{fig:recallRate}. The two
distinct co-training series are separately visualized to clearly
illustrate Recall changes at each step. We see that the SchNet0 series
somewhat plateaus in iteration `2', while the ALIGNN0 series still
improves in recall. However, neither series make significant improvement
on their recall in iteration `3'. This suggests that using the third
iteration yields diminishing returns in terms of new learning, while
risking enforcing models' biases through too many repetitions.
Furthermore, the predicted positive rate increases in both series for
iteration `3', without a meaningful increase in recall range to justify
it. This means that the model is more likely to classify a theoretical
crystal as synthesizable, without improving its understanding of
synthesizability. This is akin to over-fitting, when additional learning
steps do not yield better validation results. These factors indicate
that iteration `2' is optimal. Consequently, we omit the third iteration
and use the results from iteration `2' as the source for
synthesizability labels.

\begin{figure}[!htbp]
    \centering
    \begin{subfigure}[b]{0.45\textwidth}
        \centering
        \includegraphics[width=\textwidth]{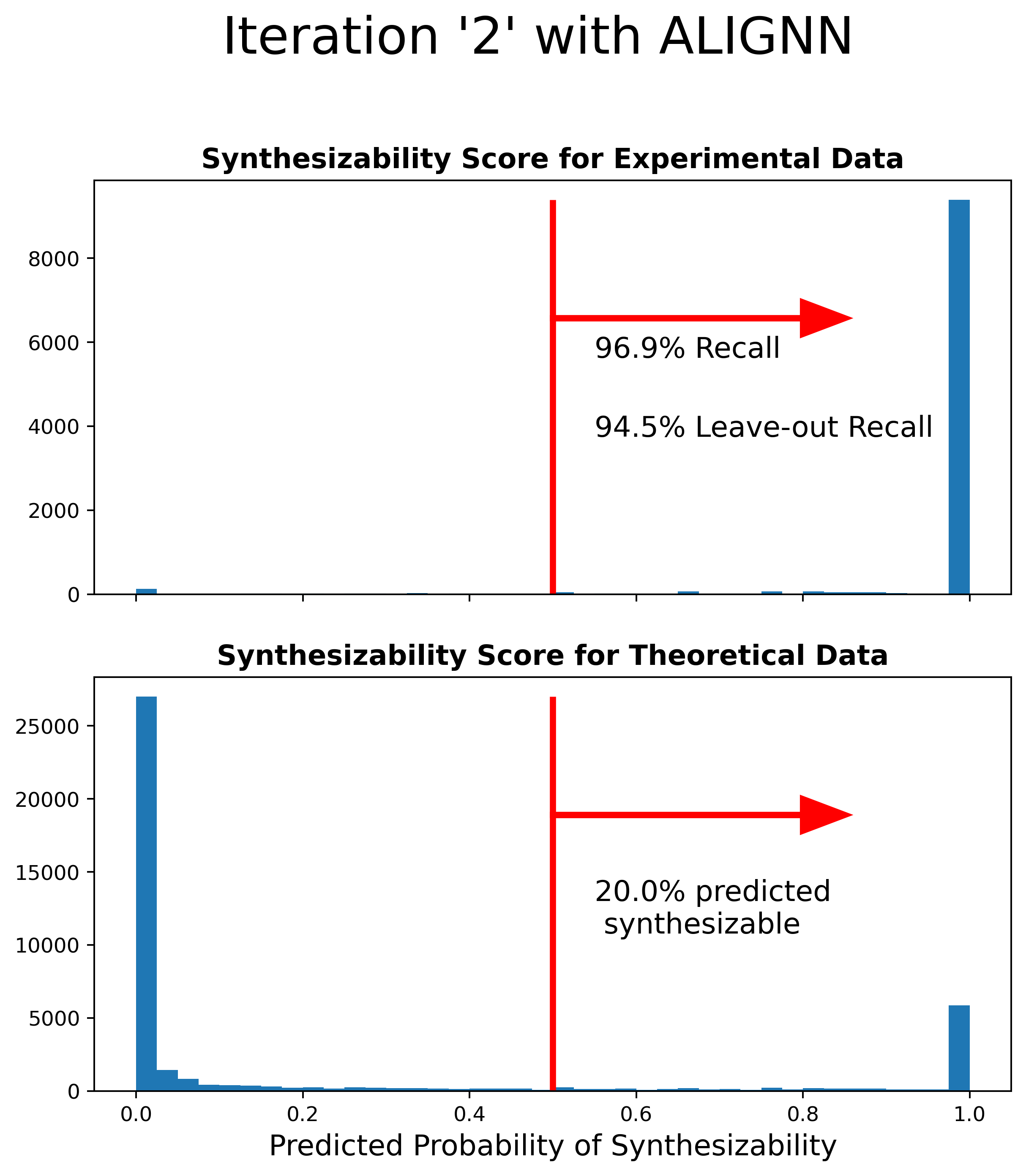}
        \caption{}
    \end{subfigure}
    \hfill
    \begin{subfigure}[b]{0.45\textwidth}
        \centering
        \includegraphics[width=\textwidth]{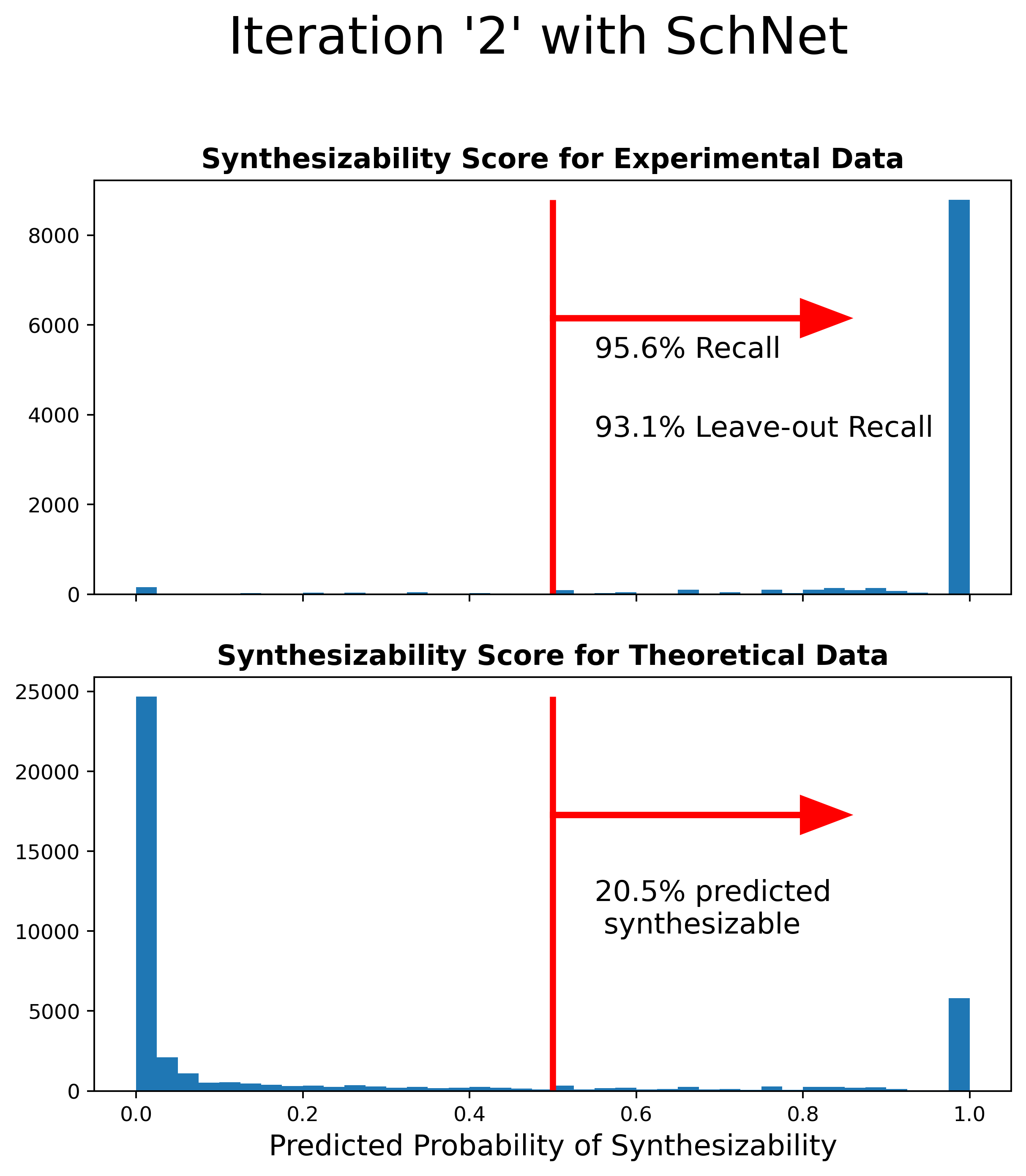}
        \caption{}
    \end{subfigure}
    \caption{\textit{Synthesizability score distribution for Iteration '2' for the a) ALIGNN0 series and b) SchNet0 series.}}
    \label{fig:synthscoredist_2}
\end{figure}

The synthesizability scores provided for the unlabeled data is the
actual goal of this PU learning task. The distribution of these scores,
alongside a large Recall rate, provide a sense of the performance
quality. A model that marks almost all crystals as synthesizable would
have a high recall but could not distinguish the two classes from each
other. Fig.~\ref{fig:synthscoredist_0} and Fig.~\ref{fig:synthscoredist_2} show the distributions of synthesizability
scores at iteration `0' and iteration `2' of co-training, respectively.
Despite high recall values, the PU learners mark only about 20\% of the
unlabeled data as synthesizable. The synthesizability scores for the
intermediate iterations can be found in the supplementary materials.

\begin{figure}[!htbp]
    \centering
        \includegraphics[width=0.8\textwidth]{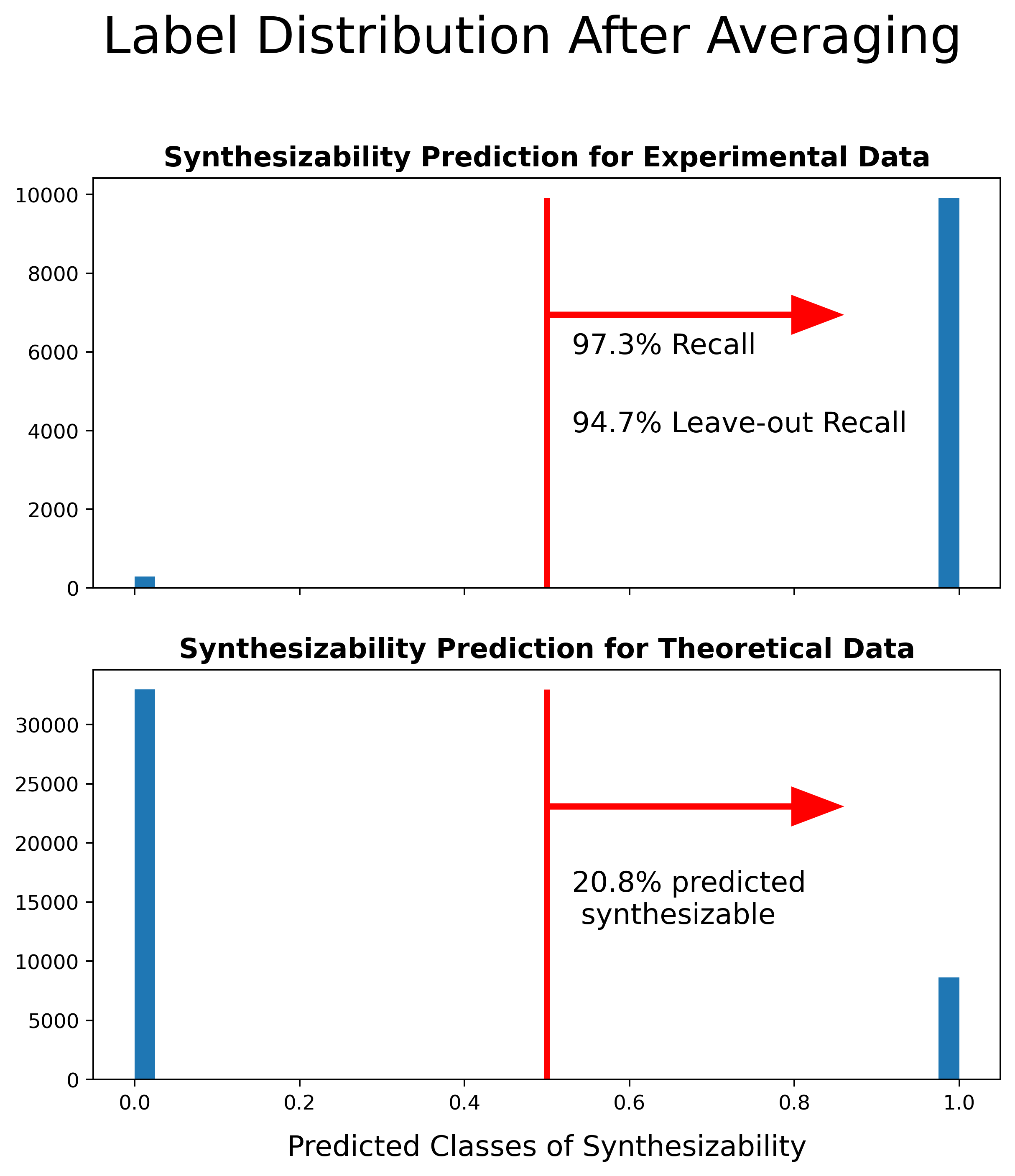}
    \caption{\textit{Label distribution after averaging scores}}
    \label{fig:finalsynthLabdist_2}
\end{figure}

In
the final step of co-training, the scores from iteration `2' are
averaged and final labels are predicted via a cutoff of 0.5. This yields
the final labels to for training the synthesizability predictor. The
recall range is now {[}95-97{]}\% and 21\% of the unlabeled data are
predicted to be synthesizable, see Fig.~\ref{fig:finalsynthLabdist_2}. Of course, all experimental
data, including the \textapprox 3\% that were misclassified as unsynthesizable, are
labeled as positive for training the synthesizability predictor.

\begin{figure}[!htbp]
    \centering
        \includegraphics[width=0.83\textwidth]{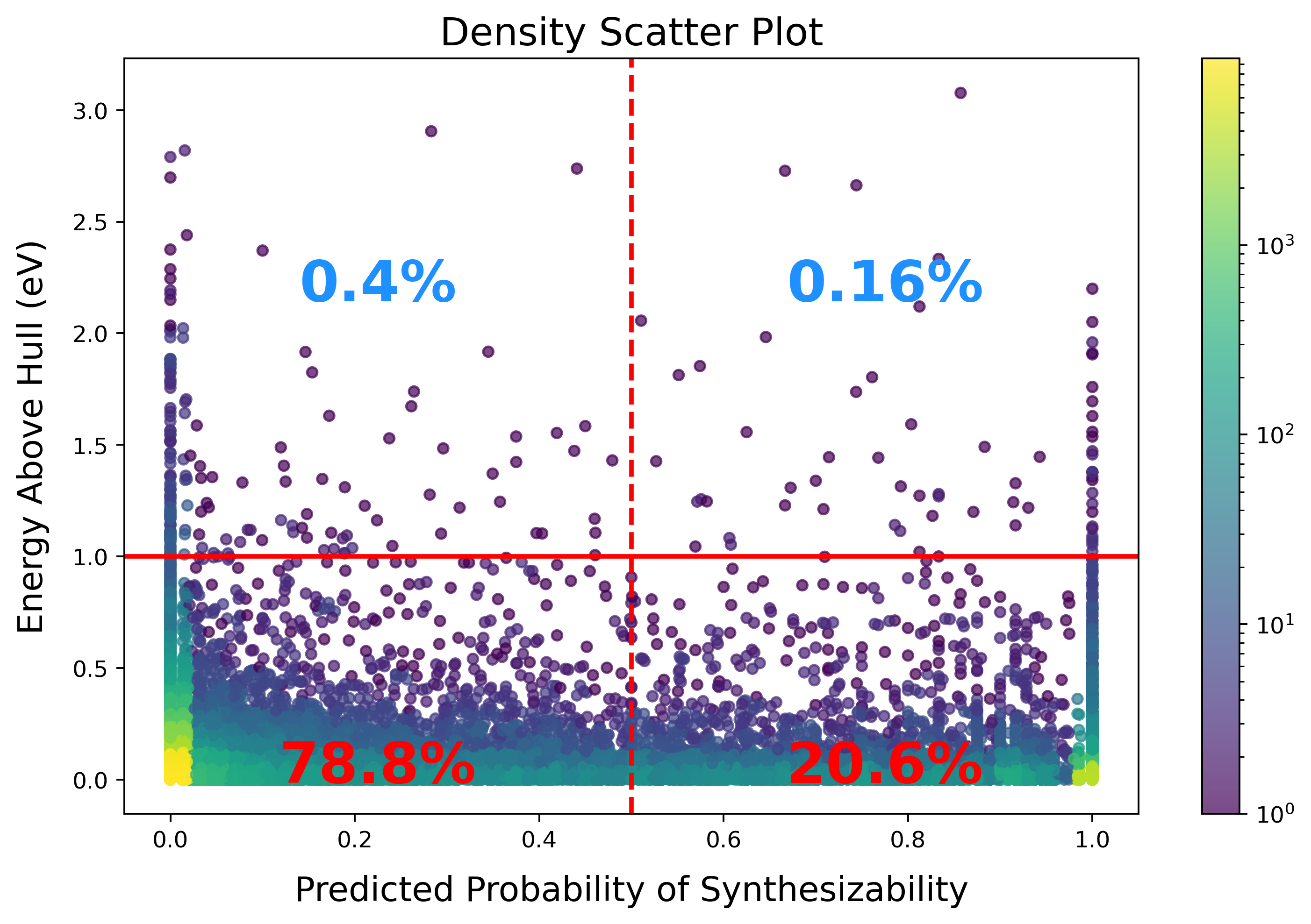}
    \caption{\textit{Density scatter plot of energy above hull vs the synthesizability score for the unlabeled data. The logarithmic color bar on the side indicates density map}}
    \label{fig:final_sctter_hm_frac_it2}
\end{figure}
Next,
we observe the synthesizability score for the unlabeled data and its
relation to stability. Crystals with energy more than 1eV above the
convex hull are quite unstable and highly unlikely to be synthesizable.
As shown in Fig.~\ref{fig:final_sctter_hm_frac_it2} these unstable crystals are 2.5 times less likely to
be classified synthesizable, as opposed to not-synthesizable. Also, more
than 99\% of the unlabeled crystals that are classified as
synthesizable, have an energy less than 1eV above the convex hull. This
shows that the model has captured the property of stability, even
without having been specifically trained on it. On the other hand, among
all the crystals with energy less than 1eV above hull, only about 21\%
are classified as synthesizable. This again indicates that while
stability is a major contributor to synthesizability, it is not
sufficient for predicting it. Of course, these observations are under
the limitations of DFT such as finite temperature effects and precision.

\begin{figure}[h]
    \centering
        \includegraphics[width=0.92\textwidth]{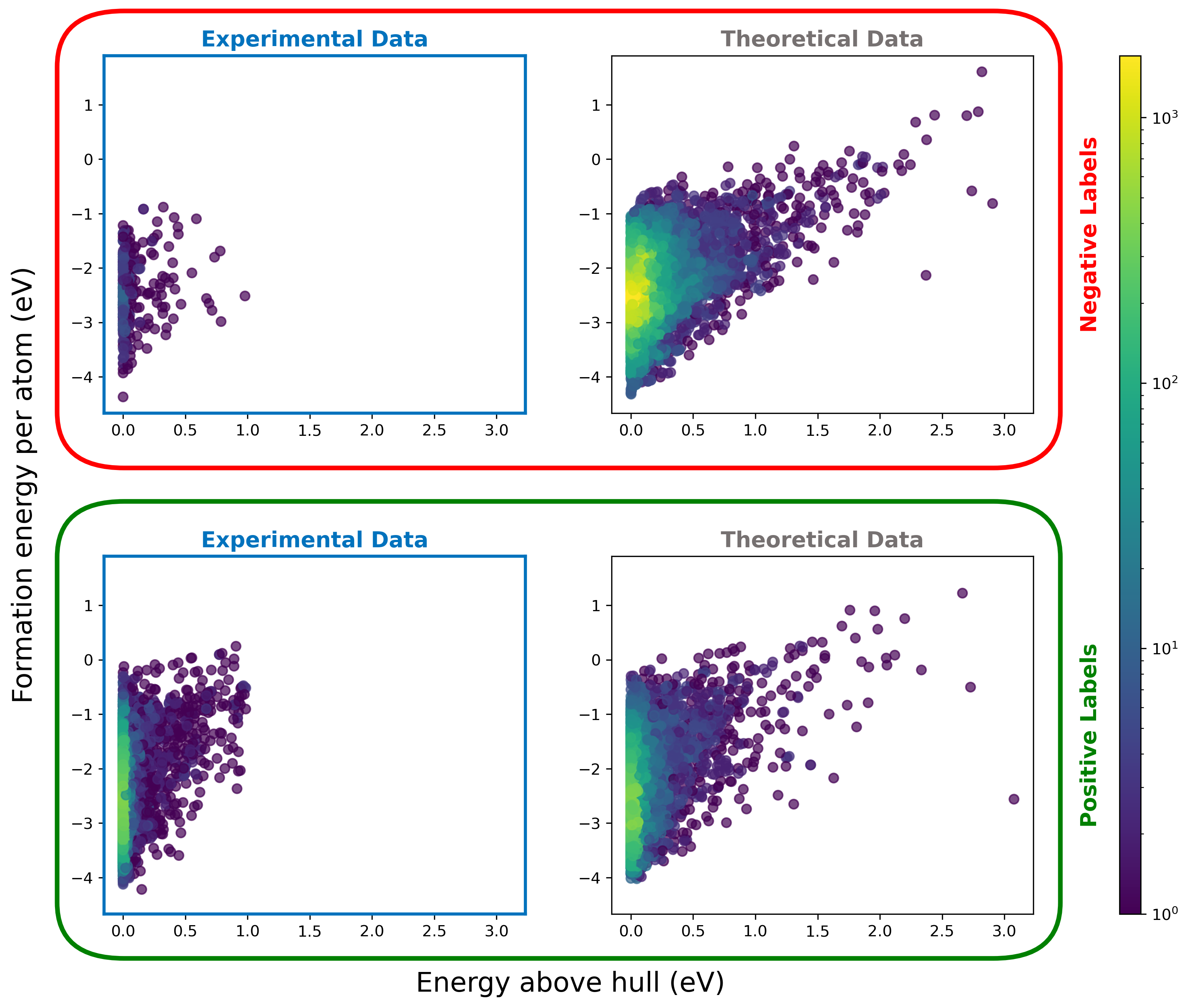}
    \caption{\textit{ Distribution of formation energy versus energy above hull for experimental data (left) and theoretical data (right), separated by predicted labels. The figure illustrates the expected clustering of positively labeled data around lower values of energy above hull while no distinct density peaks are observed with respect to formation energy. The left column validates the model’s reliability, as all experimental data are known to belong to the positive class. In contrast, the right column shows the model’s task of distinguishing between positive and negative classes within the unlabeled data.}}
    \label{fig:final_label_class_dist_it2}
\end{figure}

In Fig.~\ref{fig:final_label_class_dist_it2} we compare the energy above hull and formation energy for data
with positive and negative labels. Here, we find an expected clustering
of positive data around lower values of energy above hull without any
visible density peaks with regards to formation energy. This is
expected, as stability (and by proxy synthesizability) depend on
relative available energy states less than an absolute value. The column
on the left is a measure of how trustworthy our model has been, as we
know all the experimental data belongs to the positive class. The column
on the right demonstrates the actual task of this ensemble, which is to
separate the positive and negative classes within the unlabeled data.

\hypertarget{ground-truth-evaluation}{%
\subsection{Ground Truth evaluation}\label{ground-truth-evaluation}}

\hypertarget{test-sets-construction}{%
\subsubsection{\texorpdfstring{Test-sets construction
}{Test-sets construction }}\label{test-sets-construction}}

Construction of a suitable test-set is required to report any type error
or merit measure in machine learning; more-so when predicting
out-of-distribution data is concerned~\cite{li_critical_2023}. In their paper
establishing the bagging PU Learning method~\cite{mordelet_bagging_2010}, Mordelet and Vert
use a different test-set at each run. After all runs have been executed,
the average label predicted for each datum as part of the test-set is
taken as the probability of it belonging to the positive class. A
threshold is then applied to this probability to determine the label of
the datum, and thus, the error criteria. This does not lead to `data
leakage' as no model is tested on the data it has been trained on. This
dynamic test-set also lends itself well to co-training, as it does not
take away valuable data permanently from the growing train-sets of
further iterations.

On the other hand, using a leave-out test-set in the common practice in
materials informatics. At the very least, a leave-out test-set would
provide a more comparable evaluation with similar works in this topic.

Ultimately, the goal of a test error is to approximate the expected test
error. By using both test sets, we will have two values for Recall. That
means more information about the model's performance. We chose a
leave-out test set with 5\% of the positive data for all the runs. For
the dynamic test set, 10\% of the positive data is chosen at each run of
the PU learning.

\hypertarget{ground-truth-in-pu-learning}{%
\subsubsection{Ground truth in PU
learning}\label{ground-truth-in-pu-learning}}

Recall is the typical measure for evaluating PU learning tasks. Due to
the unlabeled data, this is not the most reliable measure. Recall only
tells us how much of the known positive samples were classified
correctly. The assumption is that the positive data are sampled from an
unknown distribution. Hence, the Recall based on the labeled data should
approximate a Recall based on all the positive data. Yet, it would be
beneficial to have some evidence, even if qualitative, that the Recall
solely based on the labeled data in fact approximates a Recall based on
all the data, the Ground Truth Recall. To that end, we construct a new
PU learning task with the goal of predicting classes of stability. As
mentioned earlier, stability and synthesizability are related
properties. If the recall values for labeled stability classes closely
approximate the Ground Truth Recall for all of the data, this suggests a
similar behavior in the recall values for synthesizability.

We use the same dataset as before, now including the outliers
of experimental structures with high energy above hull that were
previously excluded. This adjustment retains data for learning the
higher energy structure and provides a better benchmark for comparison
with previous works in synthesizability that used the outliers. These
data points were classified into positive (stable) and negative
(unstable) classes based on a cutoff in energy above the convex hull;
for details please see supplementary materials. The key difference is
that, unlike a real PU learning task, all the positive and negative
labels are available for evaluation post training. A random subset of
the positive class, with the same number of data points as the original
experimental class, kept their positive label. We then hid the label of
the remaining data to manufacture a PU learning scenario. The models
were trained on the stability PU data using the same code as the
synthesizability task. Having access to all the labels, we could
estimate the Ground Truth Recall value and compare it with the Recall
values produces by the two test sets.

\begin{figure}[!htbp]
    \centering
    \begin{subfigure}[b]{0.7\textwidth}  
        \centering
        \includegraphics[width=\textwidth]{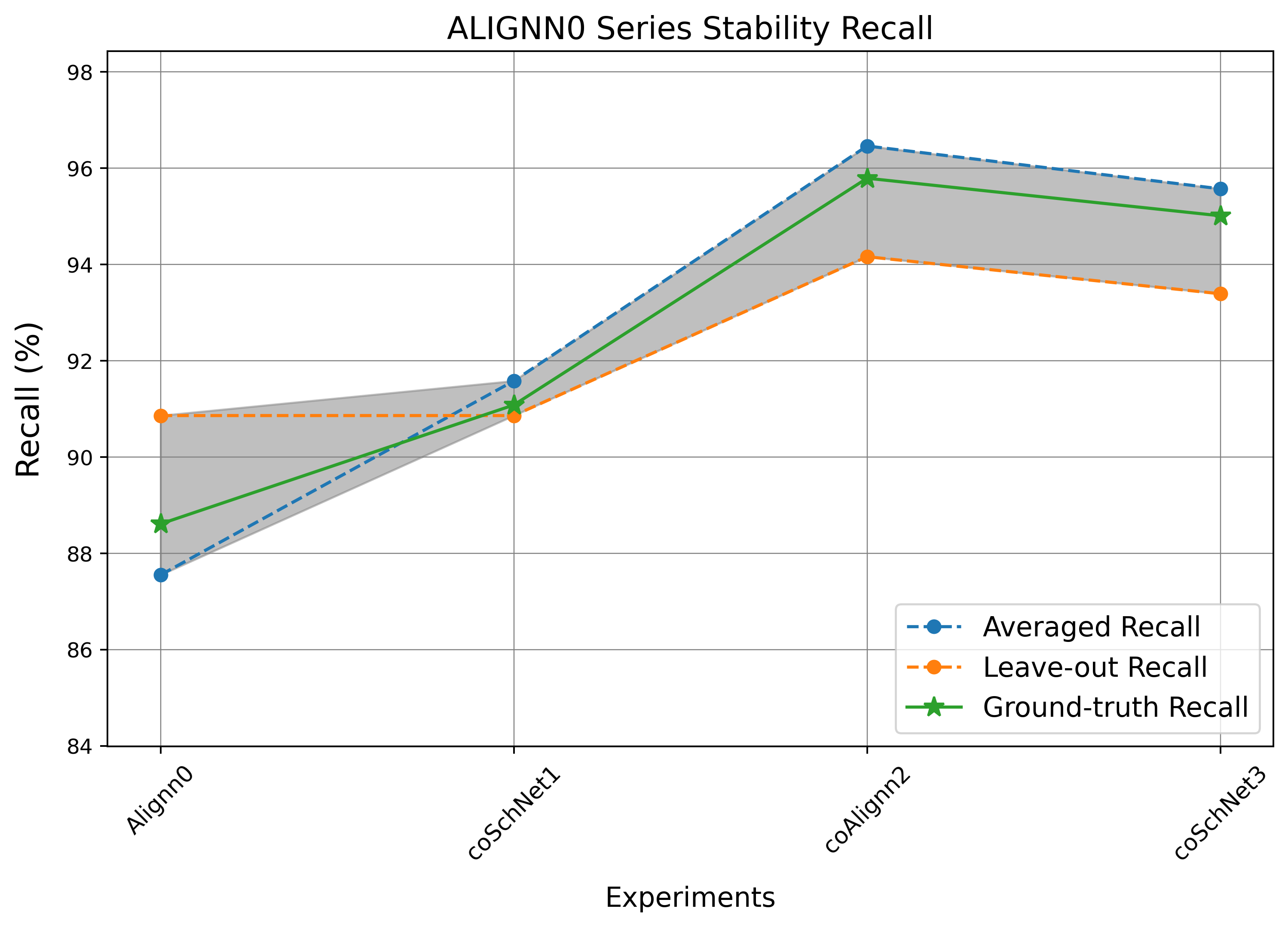} 
        \caption{}
        \label{fig:part-8a}
    \end{subfigure}
    
    \begin{subfigure}[b]{0.7\textwidth}  
        \centering
        \includegraphics[width=\textwidth]{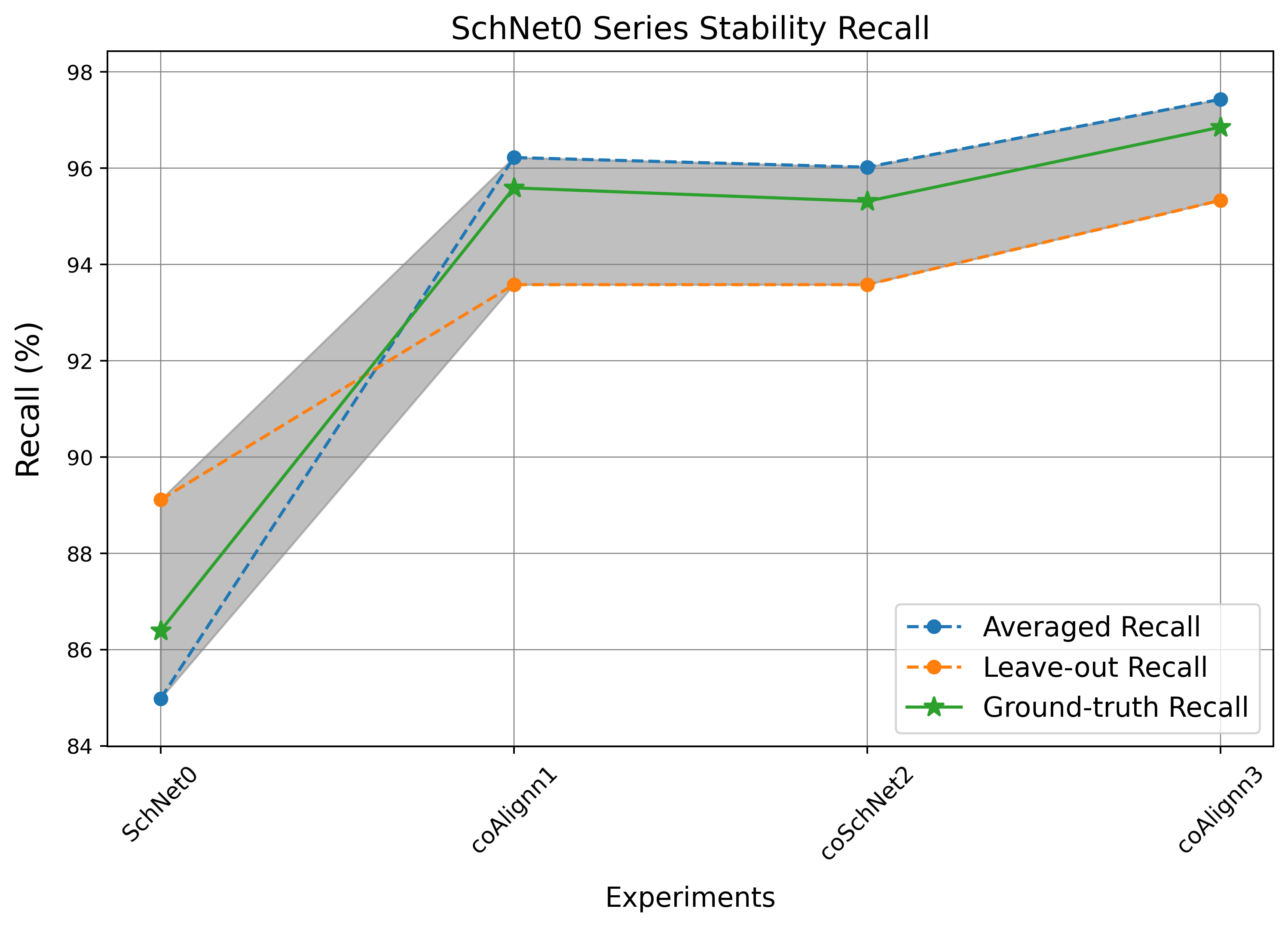}  
        \caption{}
        \label{fig:part-8b}
    \end{subfigure}
    
    \caption{\textit{ Ground Truth Recall progression per iteration for stability classe.}}
    \label{fig:StabRecallRate}
\end{figure}

As shown in Fig.~\ref{fig:StabRecallRate}, the recall values produced by both test-sets closely
approximate the Ground Truth Recall, confirming the reliability of using
Recall for evaluating the model's performance. In both co-training
series, the leave-out recall value starts more optimistic than the
ground truth, especially when high-energy experimental outliers are
included in the PU Learning. This optimistic recall was the reported
recall value in the previous PU learning studies predicting
synthesizability~\cite{jang_structure-based_2020, gu_perovskite_2022}. From iteration `1' of co-training
the order flips and the dynamic test set becomes too optimistic. While
there is no guarantee the ground-truth will always be found in the range
between the two values, Fig.~\ref{fig:StabRecallRate} illustrates why using both test-sets is
worthwhile
rather than just keeping one.

\FloatBarrier

\hypertarget{predicting-synthesizability}{%
\subsection{Predicting
Synthesizability}\label{predicting-synthesizability}}

The final step in this work involves training a synthesizability
predictor using reliable labels generated through co-training. Since
materials databases apply different criteria for data inclusion, they
exhibit varied data distributions. Consequently, achieving optimal
performance on a specific test set is insufficient. It is crucial to
avoid bias toward the Materials Project data distribution, which was
used to train the model. To mitigate this, we applied regularization
techniques during training to prevent overfitting. This was not
particularly important in the PU runs, where classifier instability
enhances the bagging process by introducing variability~\cite{mordelet_bagging_2010}. In the
final step, however, the model needs to generalize well to data
distributions not seen during training, while still maintaining good
performance on the test-set.

We selected SchNet as our classifier and achieved good results, though
other classifiers like ALIGNN can also be trained using the same labels.
Detailed training parameters are available in the METHODS section. The
pretrained model is accessible in our repository
https://github.com/BAMeScience/SynCoTrainMP).

\begin{figure}[!htbp]
    \centering
        \includegraphics[width=0.99\textwidth]{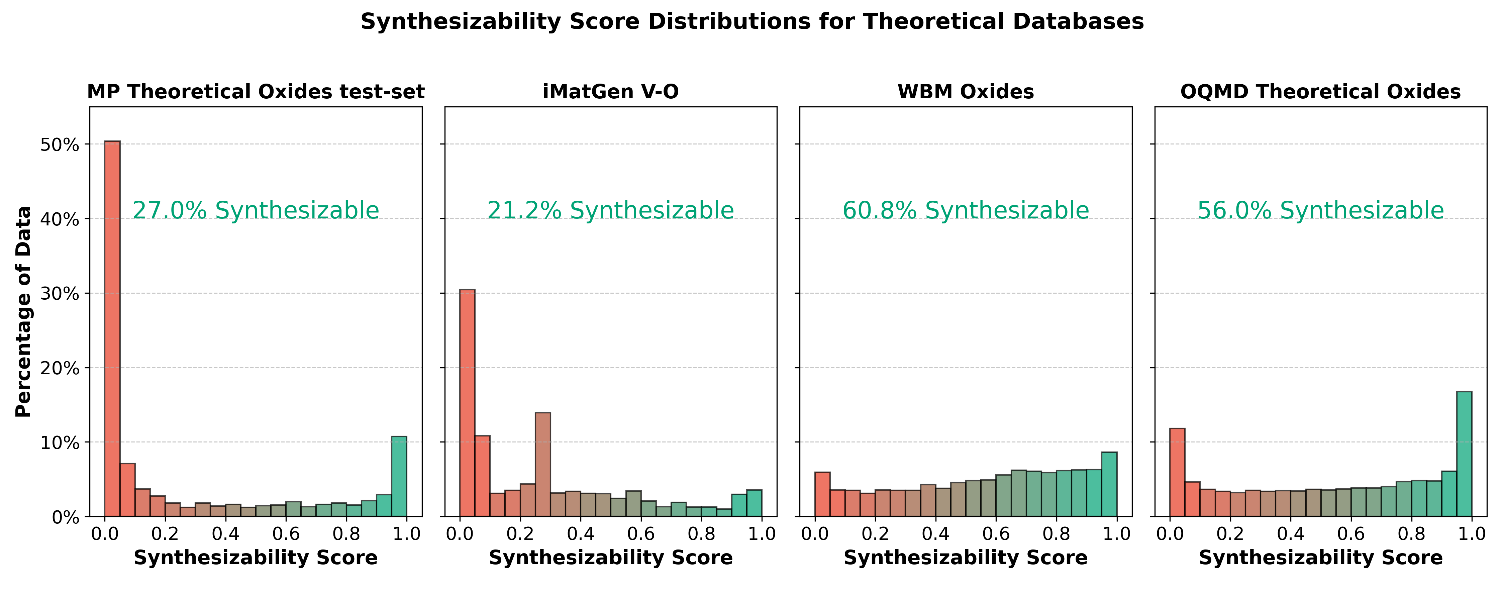}
    \caption{\textit{\textbf{Synthesizability probability distributions across theoretical databases, expressed as a percentage of each dataset.} The first distribution represents the theoretical portion of the test set, selected from Materials Project data. The second distribution corresponds to vanadium oxide crystals generated by the iMatGen generative model. The third distribution shows data from the WBM dataset, followed by the fourth, which represents oxides from the OQMD database that are absent in Materials Project experimental data.}}
    \label{fig:Synthesizability Score Distributions for Theoretical Databases}
\end{figure}

The trained model reached 90.5\% accuracy on a test set comprising 5,180
data points. To further evaluate the model's performance, we analyzed
the synthesizability predictions for three additional datasets, focusing
exclusively on oxides. First, we examined theoretical oxides from the
Open Quantum Materials Database (OQMD)~\cite{kirklin_open_2015}, downloaded via the
Jarvis Python package~\cite{choudhary_joint_2020}, after filtering out any crystals already
present in Materials Project's experimental data, leaving 23,056
theoretical oxides. Second, we analyzed 14,095 oxide crystals from the
WBM dataset~\cite{wang_predicting_2021}, which were generated using random sampling of
elements in Materials Project structures, with chemical similarity
safeguards based on ICSD data~\cite{Hellenbrandt_icsd_2004}. We used the relaxed version of
this dataset. Finally, we predicted the synthesizability of 6,156
vanadium oxide crystals generated by iMatGen~\cite{noh_inverse_2019}. Fig.~\ref{fig:Synthesizability Score Distributions for Theoretical Databases} compares
the synthesizability scores of these datasets with the theoretical
portion of the test set. All These crystal structures and their
predicted synthesizability scores are available to download in our
GitHub repository.

Over half of the theoretical test-set data shows a synthesizability
score close to zero, as expected, since previously synthesized crystals
have been excluded by Materials Project. In contrast, the OQMD data
shows roughly twice the proportion of synthesizable crystals, which may
result from differing inclusion criteria between Materials Project and
OQMD. We still observe a peak near a score of 1, possibly indicating
synthesized crystals not listed in Materials Project. The iMatGen data
show the lowest synthesizability, with multiple peaks at low scores,
reflecting the artificial nature of these generated structures, which
are often less realistic. The WBM dataset scores higher on average,
without significant peaks. Despite being artificially generated, the WBM
data employed mechanisms like chemical similarity to avoid unstable
crystals. As a result, we observe more novel crystals with ambiguous
synthesizability predictions, with scores around 0.5, and no clear peaks
close to 0 or 1.

\FloatBarrier

\hypertarget{discussion}{%
\subsection{\texorpdfstring{Discussion
}{Discussion }}\label{discussion}}

In modern material discovery, the first challenge is the abundance of
choices. The number of materials that may exist is astronomical~\cite{davies_computational_2016}
and high-throughput methods cannot screen them all. The nebulous nature
of the synthesizability question makes the development of a flawless
model challenging. The goal, however, is not perfection. Based on the
findings of this and other related studies, the majority of the
unlabeled data is determined to be unsynthesizable. Energy calculations,
while important, are not a good proxy for synthesizability. Filtering
out even half of the unsynthesizable data through synthesizability
prediction could save a significant amount of resources on simulations
and synthesis attempts. We imagine that our tool can be used in the
initial stages of materials discovery, to filter out the structures
which are not likely to result in real materials.

\begin{figure}[!ht]
    \centering
    \begin{subfigure}[b]{0.45\textwidth}
        \centering
        \includegraphics[width=\textwidth]{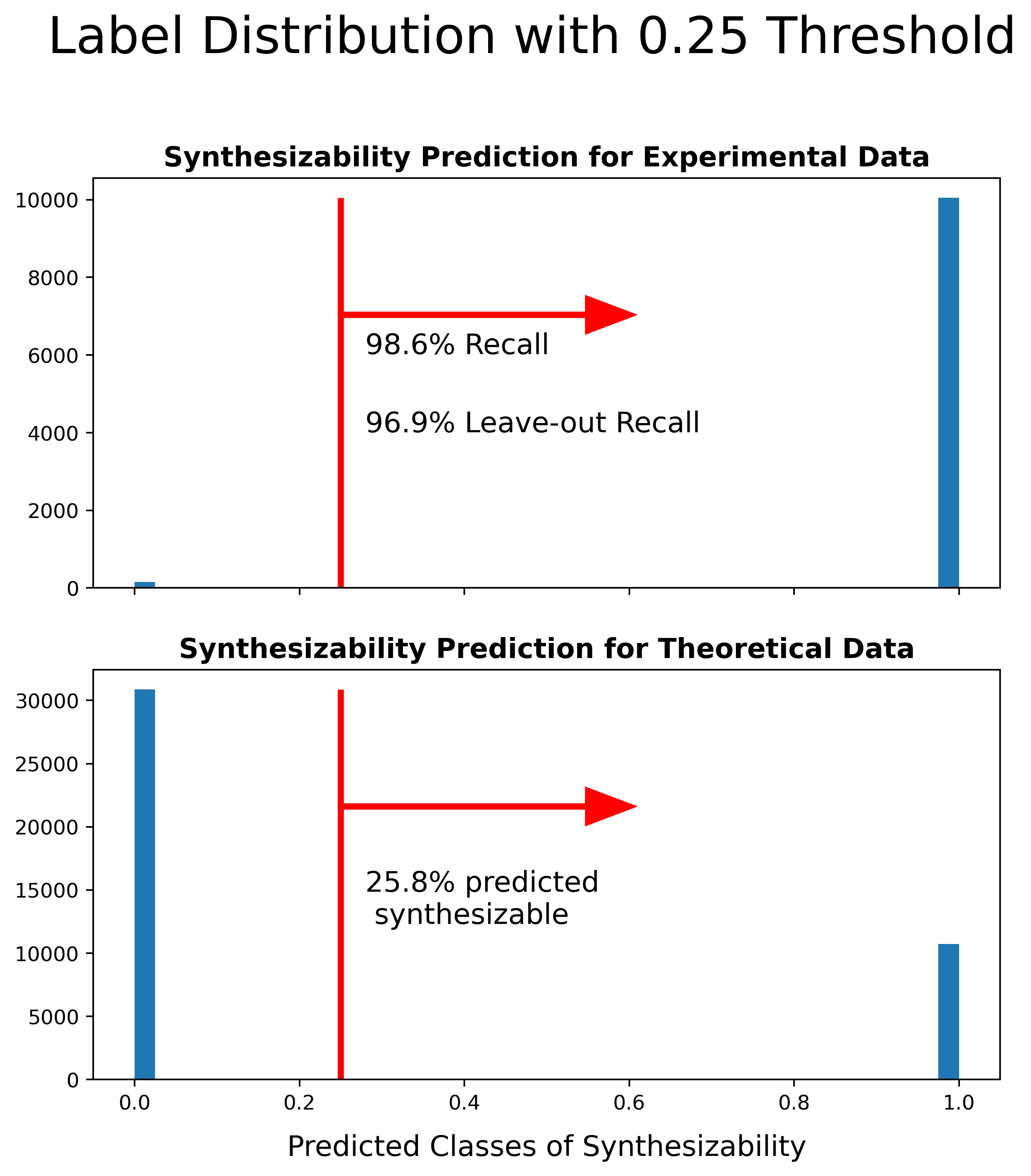}
        \caption{}
    \end{subfigure}
    \hfill
    \begin{subfigure}[b]{0.45\textwidth}
        \centering
        \includegraphics[width=\textwidth]{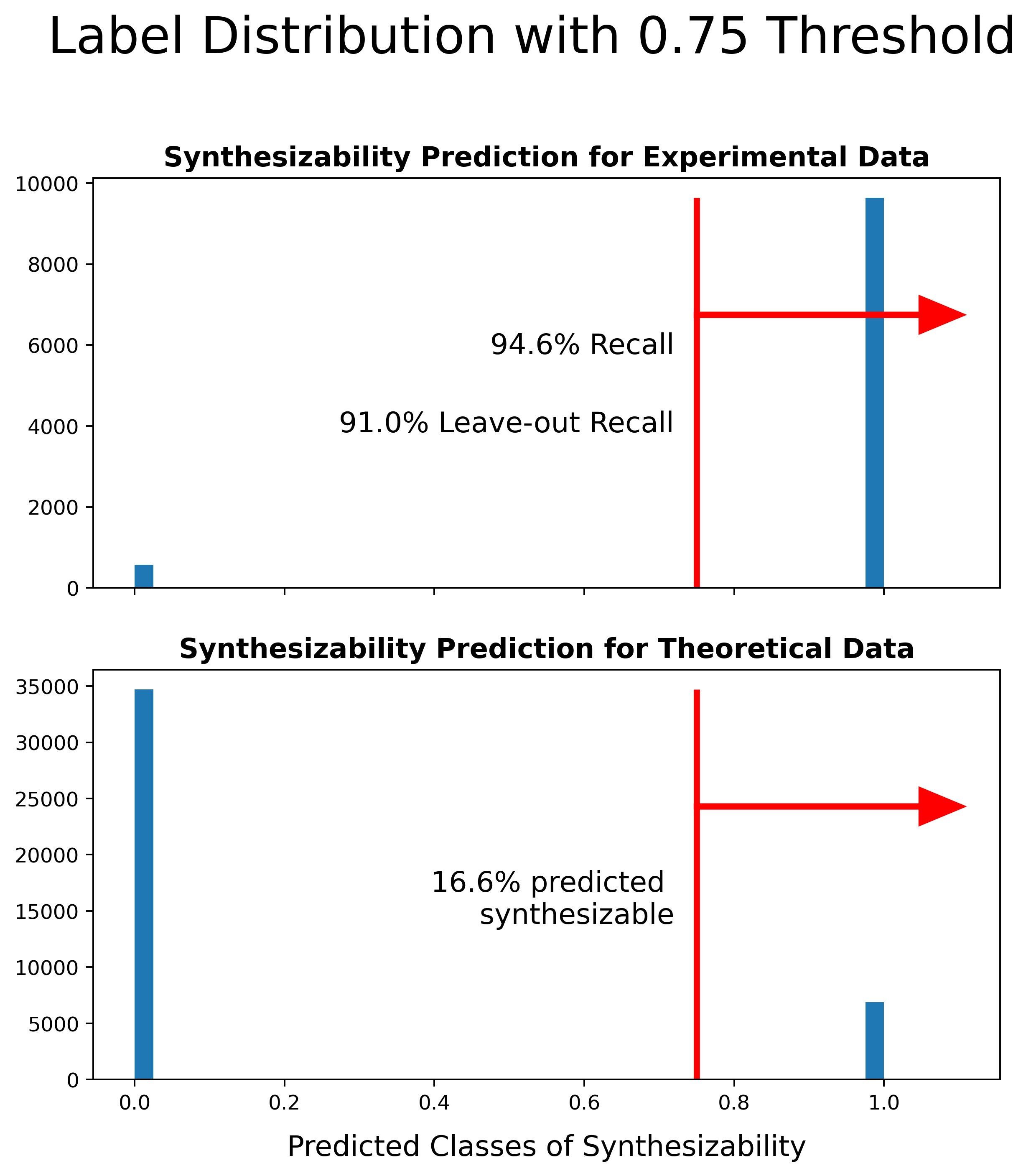}
        \caption{}
    \end{subfigure}
    \caption{\textit{Label distribution based on 0.25 and 0.75 classification thresholds at the end of co-training.}}
    \label{fig:synthscoredist_2_25_75}
\end{figure}

The decision thresholds of 0.5 and 0.75 were used as unbiased values for
classification and class expansion. However, these thresholds are
ultimately arbitrary and can be adjusted based on the specific goals and
applications. In a more exploratory study, a looser threshold could be
utilized to avoid overlooking potentially interesting novel structures.
Conversely, a project operating with a tighter budget could employ a
stricter threshold to save on resources. Label distributions based on a
threshold of 0.25 and 0.75 are illustrated as examples in Fig.~\ref{fig:synthscoredist_2_25_75}. When
compared with the unbiased threshold of 0.5 shown in Fig.~\ref{fig:finalsynthLabdist_2}, a cutoff
threshold of 0.25 is more lenient in classifying crystals as
synthesizable. However, it only identifies 26\% of the theoretical
oxides in Materials Project as synthesizable, recognizing two thirds of
the data as unsynthesizable. Conversely, a threshold of 0.75 results in
a more stringent classification, with only 17\% of the theoretical
oxides meeting this threshold. And yet, these oxides are more likely to
be synthesizable compared to those that did not meet the cut.

In this work we combined two different learners based on strong
classifiers to reach a more reliable recall. New models for predicting
materials properties are developed rapidly and materials data is
growing. Combining different tools, instead creating one from scratch,
is an untapped potential to learn more from the data already available
in the materials space.

\FloatBarrier

\hypertarget{methods}{%
\section{METHODS}\label{methods}}

\hypertarget{co-training}{%
\subsection{\texorpdfstring{Co-training
}{Co-training }}\label{co-training}}

The co-training algorithm used here was based on previous work
in~\cite{blum_combining_1998, denis_text_2003}. It is based on the idea that each data point can
be described by distinct sets of descriptors, each of which are
sufficient for learning the target. Consequently, two models learning
from different views of the data can each gain knowledge that is
inherently complementary. This is analogous to transfer learning; but
the transfer happens between the knowledge gained from different views
of the same data, rather than an auxiliary data source.

At each iteration, a base learner calculates a synthesizability score
between 0 and 1 for both the unlabeled and experimental test data. To
expand the positive class, unlabeled data points confidently classified
as positive by the PU learner are selected. Here, we use a threshold of
0.75, rather than 0.5, to determine which unlabeled data points are
added in the original positive class. After iteration `2', the scores
from both training series are averaged. The 0.5 cutoff determines the
final label.

The base learner was changed from the original naïve Bayes classifier to
a base PU learners equipped with convolutional neural networks. The
different views of the data were achieved through the difference between
the data encoding in the classifiers, i.e., ALIGNN and SchNet. Two
parallel co-training series with altering classifiers were carried out
accordingly.

\hypertarget{pu-learning}{%
\subsection{PU Learning}\label{pu-learning}}

The algorithm of PU learning was established by Mordelet and
Vert~\cite{mordelet_bagging_2010}. This method treats the unlabeled data as negative data,
contaminated with positive data. PU learning performs best when this
contamination is low.

In this work, two base PU learners were made through using two
classifiers. In both cases, a complete bagging of PU learning took 60
runs. Note that the separate runs of PU learning are not referred to as
iterations as each run is independent of the rest. This is not the case
in co-training, where each iteration depends on the results produced by
the previous iteration.

The training data at each PU learning run has a 1:1 ratio of positive
and negative labels. The size of the training set increases after each
co-training iteration, due to the expansion of the positive class. Each
run of PU learning predicts a label, 0 or 1, for the data points that
did not take part in the training phase of that run. After the 60 runs,
these predictions are averaged for each data to produce the
synthesizability score. This score is also referred to as the predicted
probability of synthesizability. The cutoff thresholds of 0.5 and 0.75
are used to predict the labels and expand the positive class,
respectively.

\hypertarget{neural-networks-architecture}{%
\subsection{Neural Networks
architecture}\label{neural-networks-architecture}}

The ALIGNN model was used according to instructions provided in its
repository. We used the version 2023.10.01 of ALIGNN.

The SchNetPack model was originally designed for regression. To
accommodate classification, a sigmoid non-linearity and a cutoff
function were added to the final layer. We used the version 1.0.0.dev0
of this model.

\hypertarget{the-synthesizability-predictor}{%
\subsection{The synthesizability predictor}\label{the-synthesizability-predictor}}

The data labeling process utilized the averaged synthesizability scores produced in iteration '2' of co-training. A cutoff of 0.75 was applied for assigning positive labels, similar to the class expansion strategy, reducing the likelihood of training on uncertain labels.

During initial tests, the predictor displayed a tendency to overestimate the positive class, likely due to overfitting to the data distribution in the Materials Project. To mitigate this, several regularization steps were introduced. First, noise was added to the labels by randomly selecting 5\% of the positive class and flipping their labels from 1 to 0. An equal number of negative class labels were also flipped from 0 to 1. This small amount of label noise helps regularize the model, preventing classifier's overconfidence in any class distribution.

Data augmentation was then employed, following a previously published method that showed significant improvements in predicting material properties. This approach perturbs atomic positions using Gaussian noise to generate slightly altered versions of the original data, which are used alongside the unperturbed data for training. This augmentation doubles the size of the training set.

The SchNet model was used as the primary synthesizability predictor, with additional regularization techniques enhancing its generalizability. A weighted loss function was employed, with a ratio of 0.45:0.55 for the positive and negative classes, respectively. This adjustment subtly discouraged over-prediction of the positive class, while maintaining model sensitivity.

Finally, to implement regularization during training, dropout layers were added to the model, with 10\% dropout at the embedding layer and 20\% at each convolutional layer. To manage the learning rate, a `Cosine Annealing with Warm Restarts` scheduler was used, allowing it to cycle through phases, helping the model escape local minima early in training while converging effectively later on. Early stopping was also implemented to prevent overtraining.

\hypertarget{datasets}{%
\subsection{Datasets}\label{datasets}}

The experimental and theoretical data for co-training was queried from the \textbf{Materials Project API}~\cite{jain_commentary_2013}, database version 2023.11.1.

\textbf{Open Quantum Materials Database (OQMD)}~\cite{kirklin_open_2015} served as an external dataset that was not used in model training.
However, they were downloaded through the \textbf{Jarvis Python package}~\cite{choudhary_joint_2020} on 2023.12.12, which provides easy access to this data.

The WBM dataset~\cite{wang_predicting_2021} was made available through the Matbench Discovery~\cite{riebesell_preprint_2023} project through figshare~\cite{noauthor_matbench_2023}.

\hypertarget{data-availability}{%
\section{DATA AVAILABILITY}\label{data-availability}}

The code for downloading the data used in this project, and the cleaned
version of the data are available at
\url{https://github.com/BAMeScience/SynCoTrainMP}.

\hypertarget{code-availability}{%
\section{CODE AVAILABILITY}\label{code-availability}}

The code used for training SynCoTrain and predicting synthesizability is
available at \url{https://github.com/BAMeScience/SynCoTrainMP}.

\bibliography{synthcotrain}

\end{document}


\section{Supplemental Material}

\subsection{Synthesizability score distribution for all iterations}

\begin{figure}[!htbp]
    \centering
    \begin{subfigure}[b]{0.45\textwidth}
        \centering
        \includegraphics[width=\textwidth]{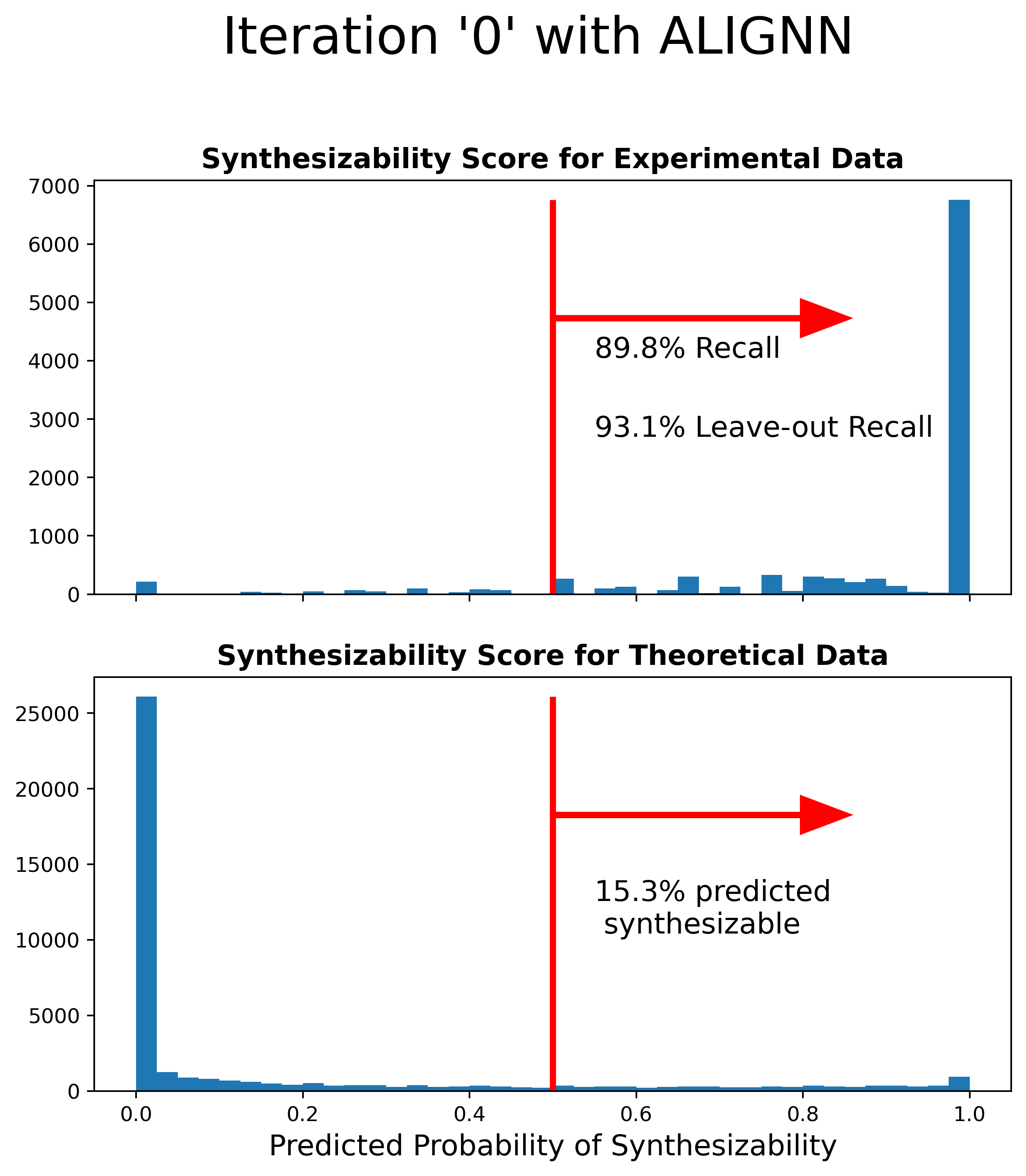}
        \caption{}
    \end{subfigure}
    \hfill
    \begin{subfigure}[b]{0.45\textwidth}
        \centering
        \includegraphics[width=\textwidth]{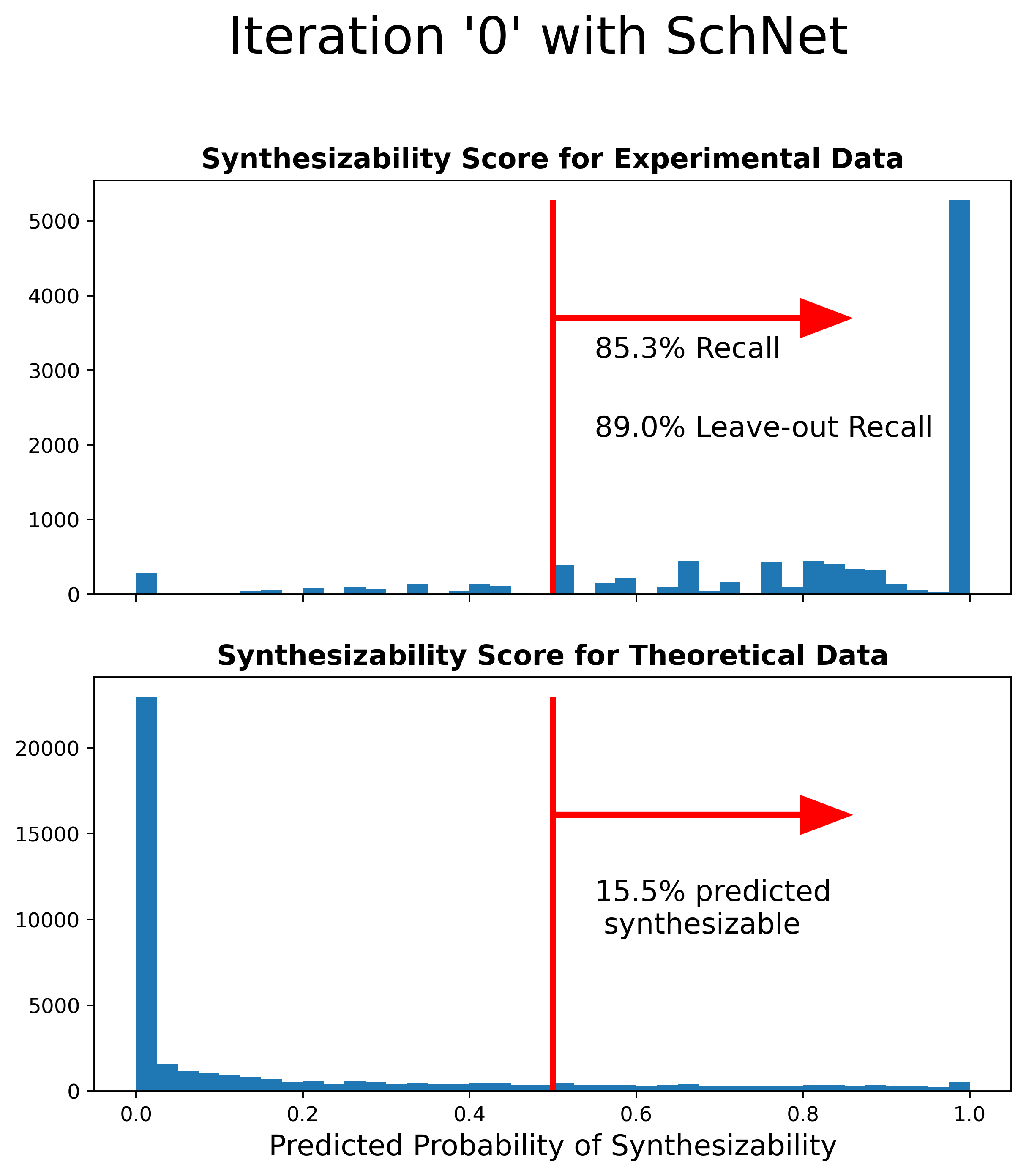}
        \caption{}
    \end{subfigure}
    \caption{\textit{Synthesizability score distribution for Iteration '0' for the a) ALIGNN0 series and b) SchNet0 series.}}
    \label{fig:synthscoredist_0}
\end{figure}
\textbf{\hfill\break
}

\begin{figure}[!htbp]
    \centering
    \begin{subfigure}[b]{0.45\textwidth}
        \centering
        \includegraphics[width=\textwidth]{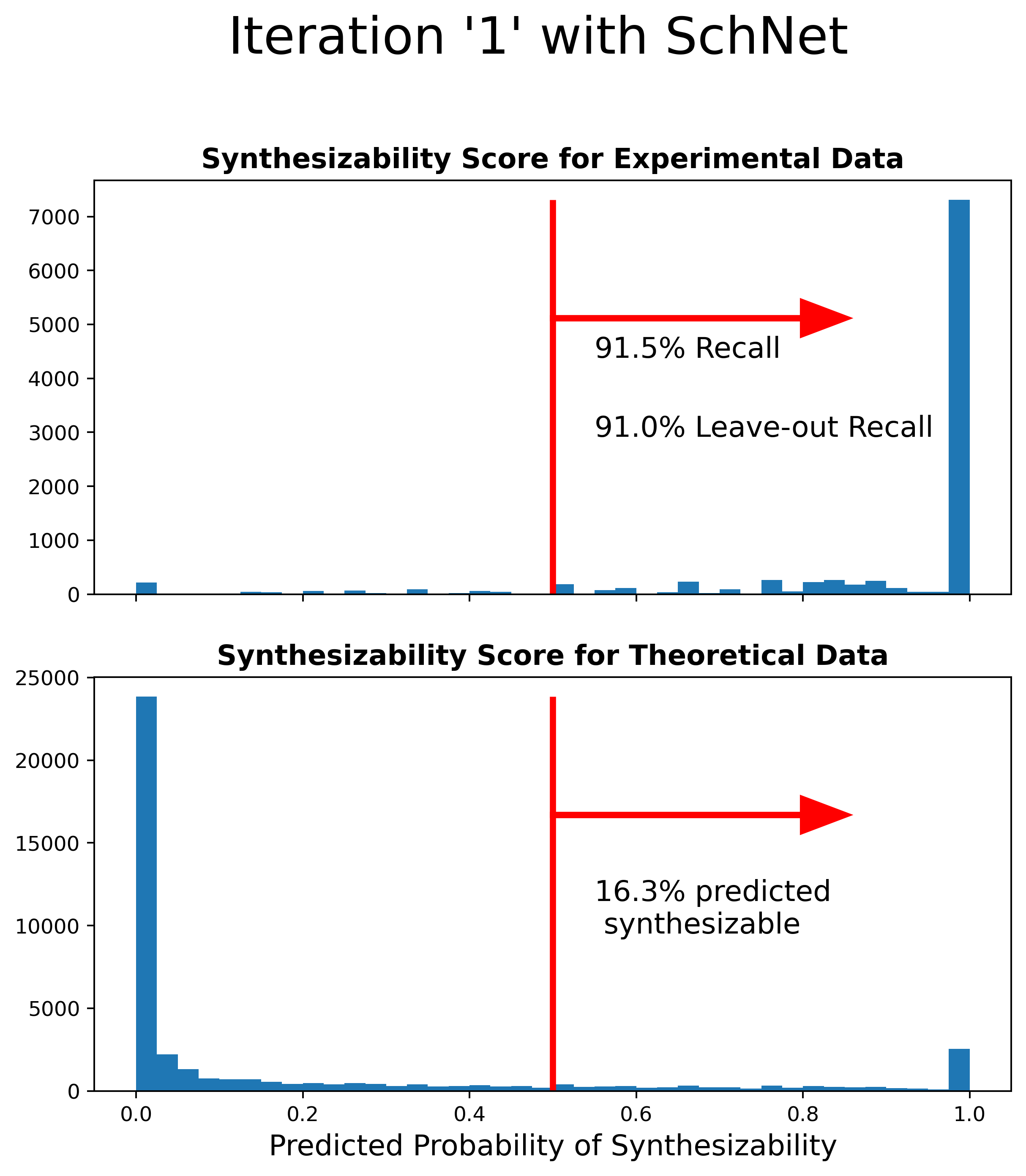}
        \caption{}
    \end{subfigure}
    \hfill
    \begin{subfigure}[b]{0.45\textwidth}
        \centering
        \includegraphics[width=\textwidth]{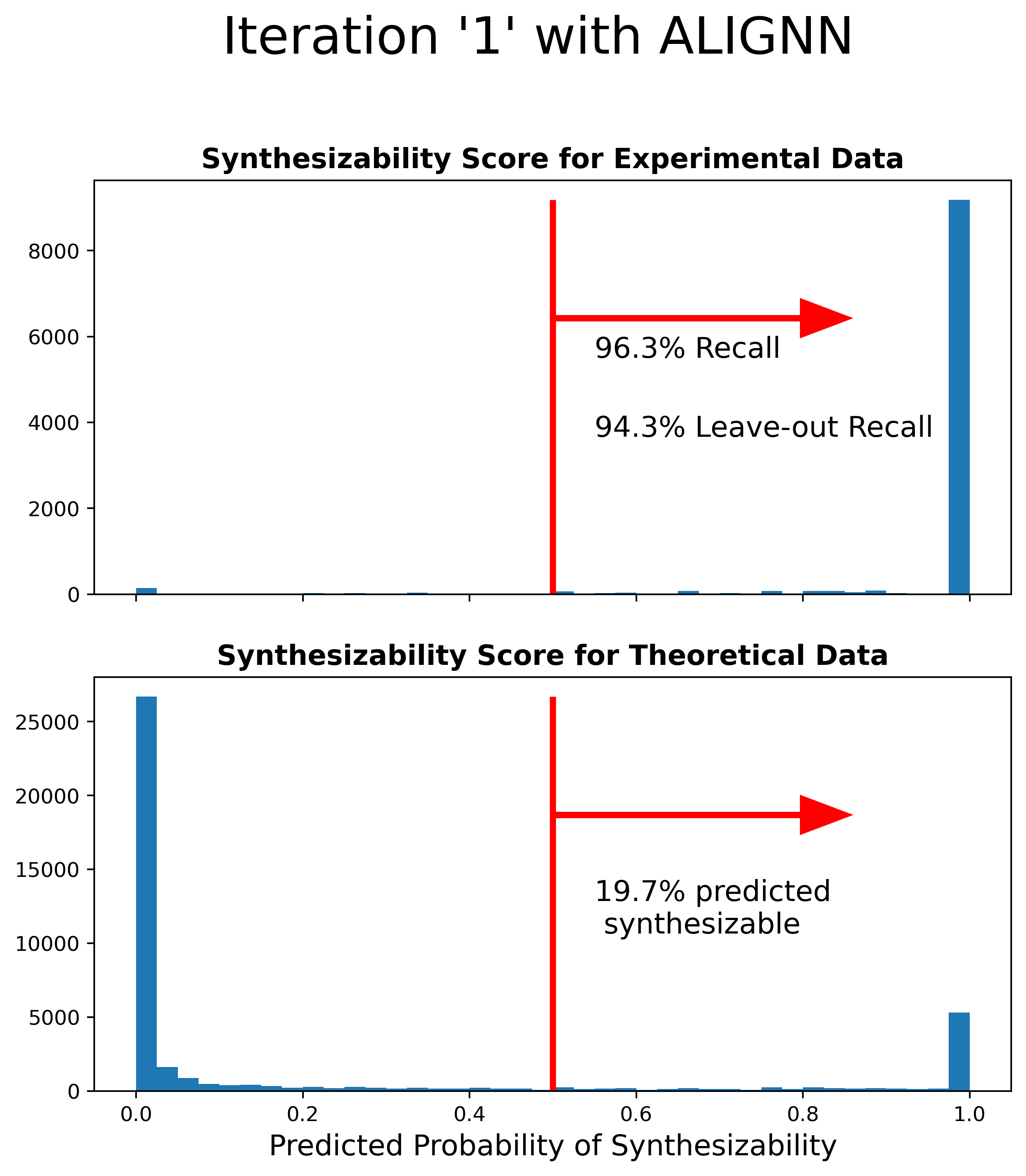}
        \caption{}
    \end{subfigure}
    \caption{\textit{Synthesizability score distribution for Iteration '1' for the a) ALIGNN0 series and b) SchNet0 series.}}
    \label{fig:synthscoredist_1}
\end{figure}

\begin{figure}[!htbp]
    \centering
    \begin{subfigure}[b]{0.45\textwidth}
        \centering
        \includegraphics[width=\textwidth]{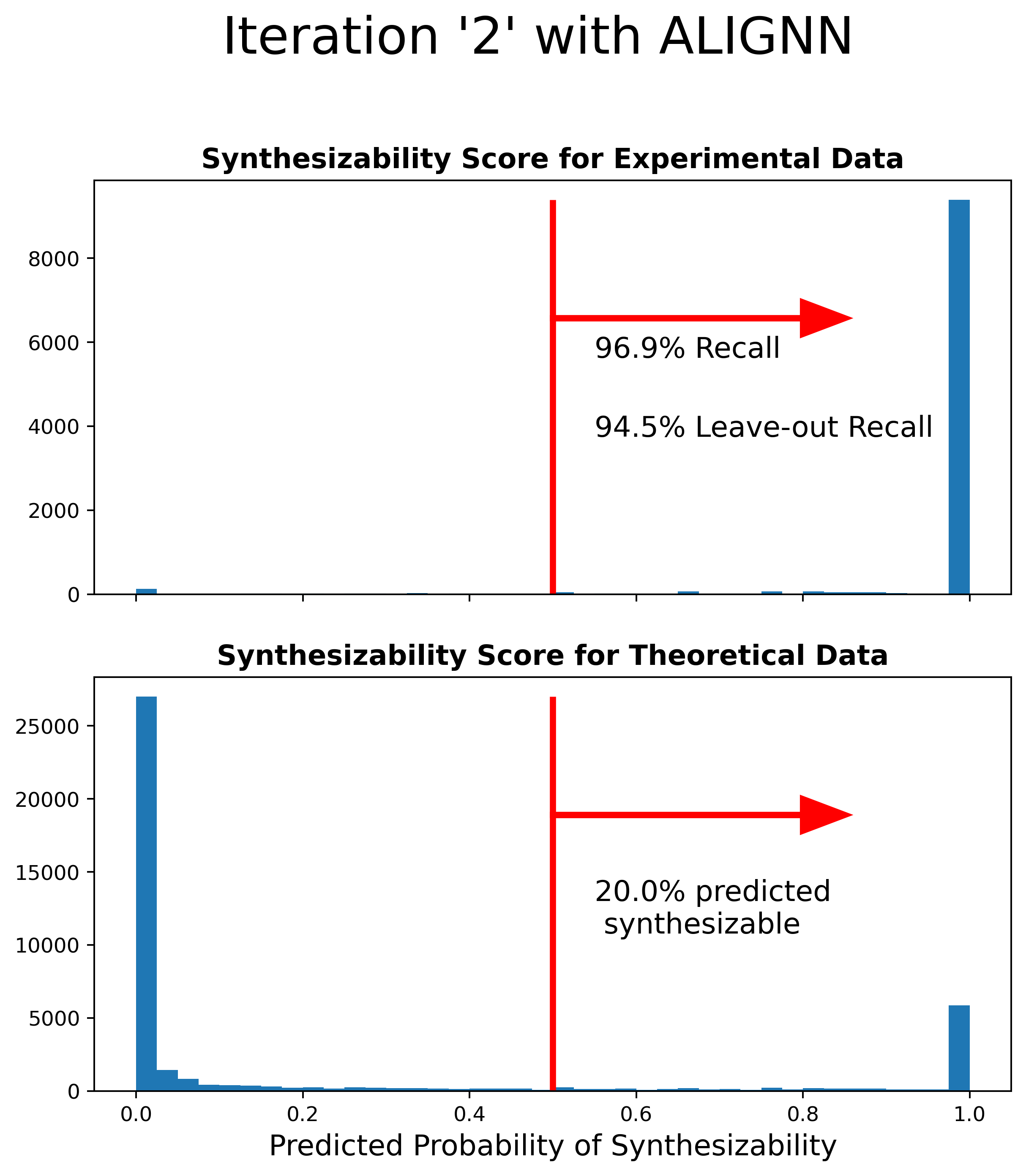}
        \caption{}
    \end{subfigure}
    \hfill
    \begin{subfigure}[b]{0.45\textwidth}
        \centering
        \includegraphics[width=\textwidth]{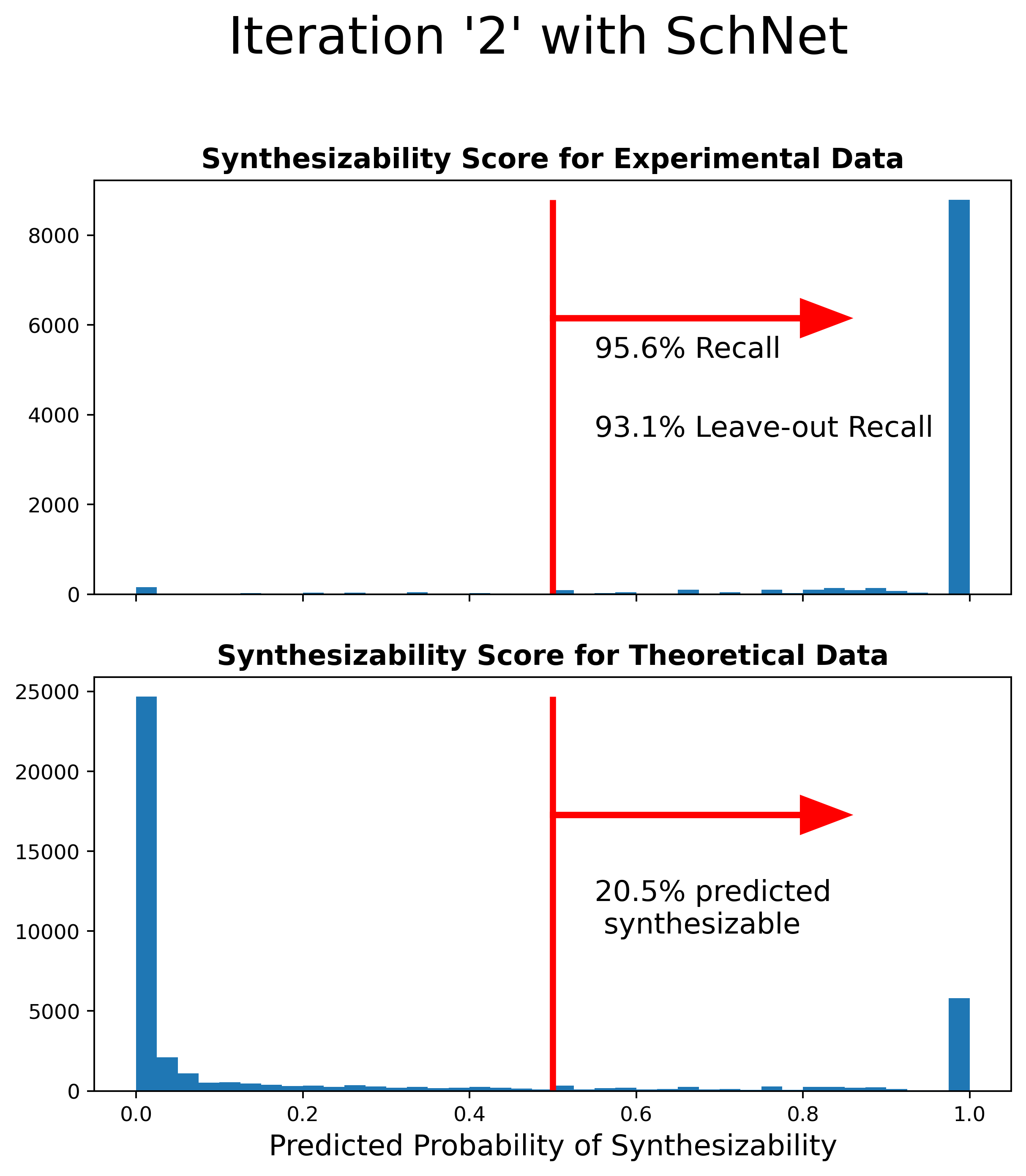}
        \caption{}
    \end{subfigure}
    \caption{\textit{Synthesizability score distribution for Iteration '2' for the a) ALIGNN0 series and b) SchNet0 series.}}
    \label{fig:synthscoredist_2}
\end{figure}

\begin{figure}[!htbp]
    \centering
    \begin{subfigure}[b]{0.45\textwidth}
        \centering
        \includegraphics[width=\textwidth]{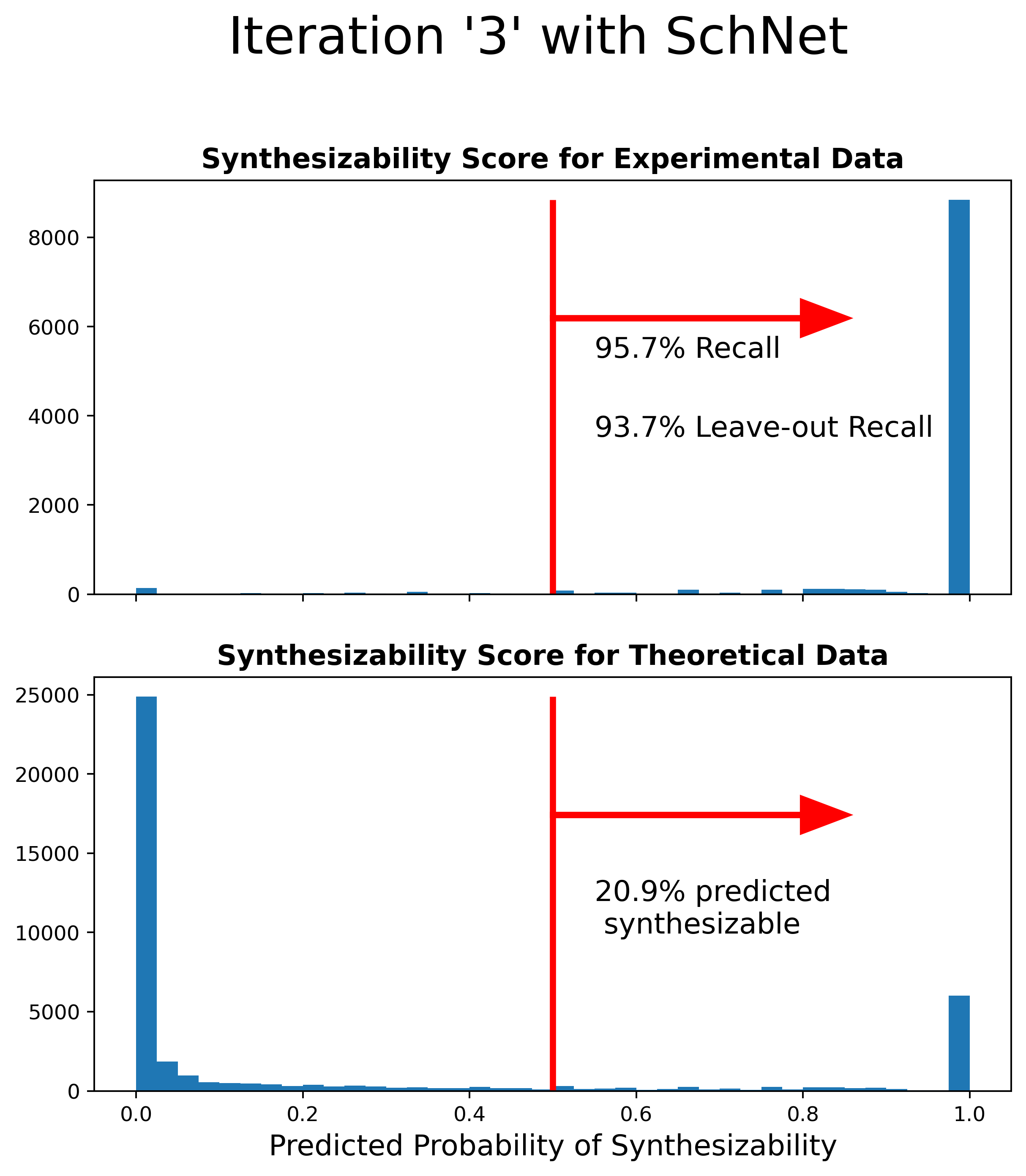}
        \caption{}
    \end{subfigure}
    \hfill
    \begin{subfigure}[b]{0.45\textwidth}
        \centering
        \includegraphics[width=\textwidth]{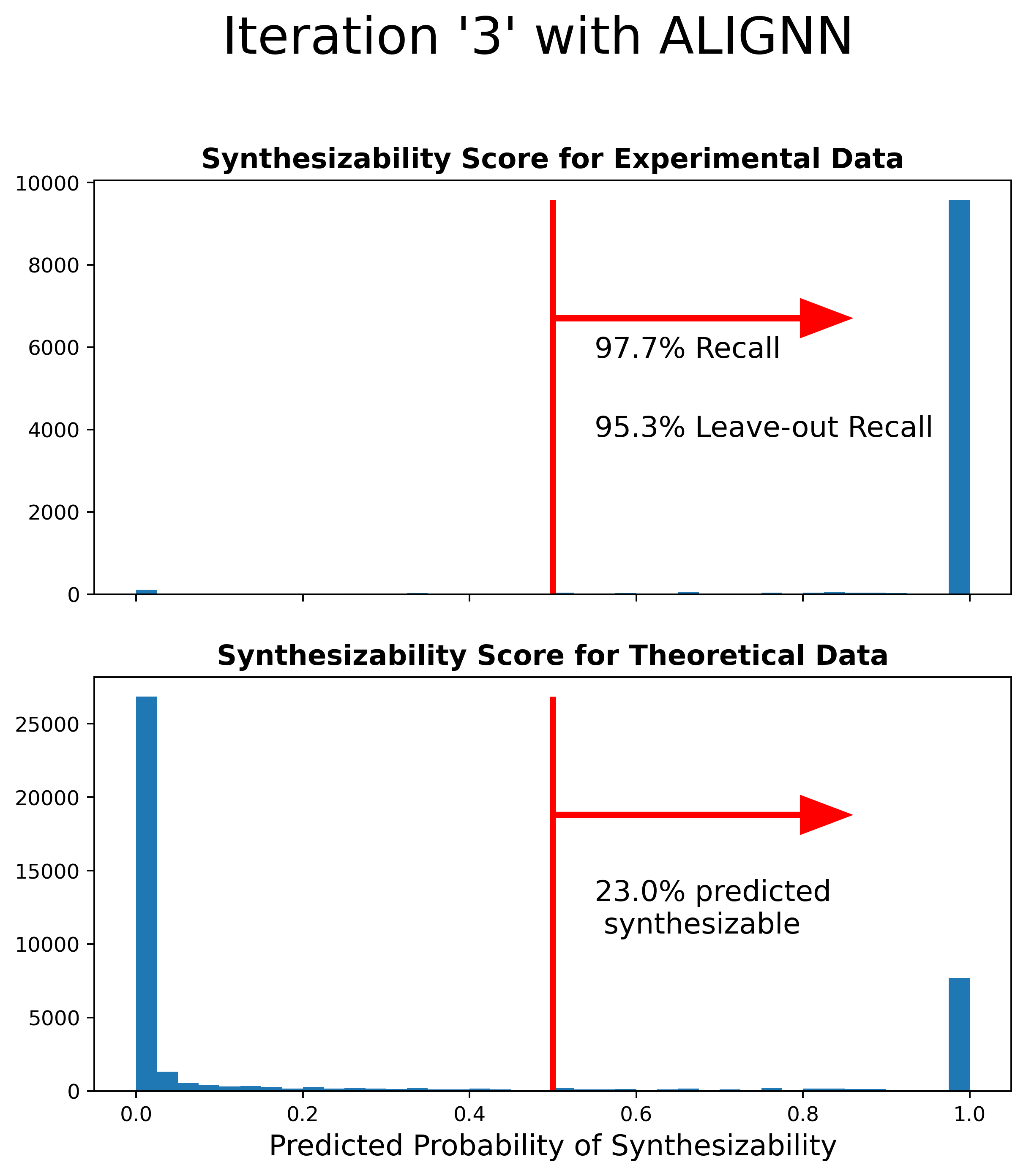}
        \caption{}
    \end{subfigure}
    \caption{\textit{Synthesizability score distribution for Iteration '3' for the a) ALIGNN0 series and b) SchNet0 series.}}
    \label{fig:synthscoredist_3}
\end{figure}

\subsection{Ground truth stability set-up}
This auxiliary experiment does not affect the synthesizability prediction results but serves as a demonstration of the reliability of the Recall values. To set up this experiment, we needed a threshold for the energy above the convex hull, which would allow us to label materials as stable (1) or unstable (0). While the obvious choice would have been 0.1 eV, a commonly used threshold for stability, it would not work well for our demonstration.

The performance of the bagging PU Learning algorithm depends on the proportion of unlabeled data belonging to the positive class—what Mordelet and Vert referred to as contamination.\cite{mordelet_bagging_2010} The algorithm assumes that part of the unlabeled data is from the negative class, and the smaller the contamination, the more accurate this assumption becomes, resulting in better performance. However, about 72\% of our data have an energy above the hull of 0.1 eV or less. If we used 0.1 eV as the threshold, the Recall values, while consistent with the ground truth, would be quite low—making it a poor demonstration of the algorithm’s abilities.
Instead, we chose 0.015 eV as the threshold for stability, which labels approximately a quarter (26\%) of our data as stable. This proportion aligns better with what we expect for synthesizability and provides a more suitable demonstration of the model’s capabilities. 

There is a valid concern that with such a low threshold, we may be approaching the precision limits of DFT, potentially producing unreliable labels. While this would be a legitimate issue if the goal were to predict stability accurately, we argue that this threshold remains useful for our purposes. Neural networks are highly capable models, able to learn any target, including the imperfections of computational models—much like what is achieved in $\Delta$-Machine Learning.\cite{ramakrishnan_big_2015}
 Therefore, it is worth evaluating our results to see if the model has, in fact, learned stability as a property.

As with synthesizability, stability labels were generated from iteration ‘2’. The label accuracy was 87\%, with 6602 crystals misclassified compared to the ground truth label produced by the 0.015 eV threshold. Of these 6602, 97\% were cases where the prediction was 1 (stable) but the actual label was 0 (unstable). Upon closer inspection, 85\% of these misclassified data points had an energy above the hull below 0.1 eV, compared to 72\% for the full dataset. This indicates that the model has indeed learned to associate lower energy above the hull with stability, even when it makes mistakes.
To verify this further, a Z-test was conducted to compare whether data points misclassified as label 1 had a significantly higher proportion of materials with energy above the hull below 0.1 eV compared to the rest of the dataset. The p-value was highly significant (p < 0.0001).

This result strongly suggests that the model has learned stability as a property, even if our chosen threshold differs from the conventional one.

\textbf{\hfill\break
}

\bibliography{synthcotrain}